\documentclass[%
reprint,
superscriptaddress,
preprintnumbers,
 amsmath,amssymb,
 aps,prc
]{revtex4-1} %arxiv

\usepackage{graphicx}% Include figure files
\usepackage{dcolumn}% Align table columns on decimal point
\usepackage{bm}% bold math

\usepackage{hyperref}% add hypertext capabilities
\usepackage{url}
\usepackage{multirow}
\usepackage{xfrac}
\begin{document}

%\preprint{APS/123-QED}

\title{Decay studies of the long-lived states in $^{186}$Tl}
\author{M.~Stryjczyk}
\email{marek.stryjczyk@kuleuven.be}
\affiliation{KU Leuven, Instituut voor Kern- en Stralingsfysica, Celestijnenlaan 200D, 3001 Leuven, Belgium}
\author{B.~Andel}
\affiliation{KU Leuven, Instituut voor Kern- en Stralingsfysica, Celestijnenlaan 200D, 3001 Leuven, Belgium}
\affiliation{Department of Nuclear Physics and Biophysics, Comenius University in Bratislava, 84248 Bratislava, Slovakia}
\author{A.~N.~Andreyev} %ICC
\affiliation{Department of Physics, University of York, YO10 5DD, York, UK}
\affiliation{Advanced Science Research Center, Japan Atomic Energy Agency, Tokai-mura, Japan}
\author{J.~Cubiss} %ICC
\affiliation{ISOLDE, CERN, CH-1211 Geneva 23, Switzerland}
\affiliation{Department of Physics, University of York, YO10 5DD, York, UK}
\author{J.~Pakarinen}
\affiliation{University of Jyv\"{a}skyl\"{a}, Department of Physics, P.O. Box 35, FI-40014, Jyv\"{a}skyl\"{a}, Finland}
\affiliation{Helsinki Institute of Physics, University of Helsinki, P.O. Box 64, FIN-00014, Helsinki, Finland}
\author{K.~Rezynkina}
\affiliation{KU Leuven, Instituut voor Kern- en Stralingsfysica, Celestijnenlaan 200D, 3001 Leuven, Belgium}
\affiliation{Universit\'e de Strasbourg, CNRS, IPHC UMR7178, F-67000, Strasbourg, France}
\author{P.~Van~Duppen}
\affiliation{KU Leuven, Instituut voor Kern- en Stralingsfysica, Celestijnenlaan 200D, 3001 Leuven, Belgium}
\author{S.~Antalic}
\affiliation{Department of Nuclear Physics and Biophysics, Comenius University in Bratislava, 84248 Bratislava, Slovakia}
\author{T.~Berry}
\affiliation{Department of Physics, University of Surrey, Guildford GU2 7XH, United Kingdom}
\author{M.~J.~G.~Borge} %ICC
\affiliation{Instituto de Estructura de la Materia, CSIC, Serrano 113 bis, E-28006 Madrid, Spain}
\affiliation{ISOLDE, CERN, CH-1211 Geneva 23, Switzerland}
\author{C.~Clisu}
\affiliation{“Horia Hulubei” National Institute for Physics and Nuclear Engineering, RO-077125 Bucharest, Romania}
\author{D.~M.~Cox}
\affiliation{Department of Physics, Lund University, Lund S-22100, Sweden}
\author{H.~De~Witte} %ICC
\affiliation{KU Leuven, Instituut voor Kern- en Stralingsfysica, Celestijnenlaan 200D, 3001 Leuven, Belgium}
\author{L.~M.~Fraile}
\affiliation{Grupo de F\'isica Nuclear, Universidad Complutense de Madrid, 28040, Madrid, Spain}
\author{H.~O.~U.~Fynbo} %ICC
\affiliation{Department of Physics and Astronomy, Aarhus University, DK-8000 Aarhus C, Denmark}
\author{L.~P.~Gaffney}
\affiliation{ISOLDE, CERN, CH-1211 Geneva 23, Switzerland}
\author{L.~J.~Harkness-Brennan} %ICC
\affiliation{Department of Physics, Oliver Lodge Laboratory, University of Liverpool, Liverpool L69 7ZE, United Kingdom}
\author{M.~Huyse}
\affiliation{KU Leuven, Instituut voor Kern- en Stralingsfysica, Celestijnenlaan 200D, 3001 Leuven, Belgium}
\author{A.~Illana}
\affiliation{Istituto Nazionale di Fisica Nucleare, Laboratori Nazionali di Legnaro, Legnaro 35020, Italy}
\affiliation{University of Jyv\"{a}skyl\"{a}, Department of Physics, P.O. Box 35, FI-40014, Jyv\"{a}skyl\"{a}, Finland}
\affiliation{Helsinki Institute of Physics, University of Helsinki, P.O. Box 64, FIN-00014, Helsinki, Finland}
\author{D.~S.~Judson} %ICC
\affiliation{Department of Physics, Oliver Lodge Laboratory, University of Liverpool, Liverpool L69 7ZE, United Kingdom}
\author{J.~Konki}
\affiliation{ISOLDE, CERN, CH-1211 Geneva 23, Switzerland}
\author{J.~Kurcewicz} %ICC
\affiliation{ISOLDE, CERN, CH-1211 Geneva 23, Switzerland}
\author{I.~Lazarus} %ICC
\affiliation{STFC Daresbury, Daresbury, Warrington WA4 4AD, United Kingdom}
\author{R.~Lica}
\affiliation{“Horia Hulubei” National Institute for Physics and Nuclear Engineering, RO-077125 Bucharest, Romania}
\affiliation{ISOLDE, CERN, CH-1211 Geneva 23, Switzerland}
\author{M.~Madurga} %ICC
\affiliation{ISOLDE, CERN, CH-1211 Geneva 23, Switzerland}
\author{N.~Marginean} %ICC
\affiliation{“Horia Hulubei” National Institute for Physics and Nuclear Engineering, RO-077125 Bucharest, Romania}
\author{R.~Marginean} %ICC
\affiliation{“Horia Hulubei” National Institute for Physics and Nuclear Engineering, RO-077125 Bucharest, Romania}
\author{C.~Mihai} %ICC
\affiliation{“Horia Hulubei” National Institute for Physics and Nuclear Engineering, RO-077125 Bucharest, Romania}
\author{P.~Mosat}
\affiliation{Department of Nuclear Physics and Biophysics, Comenius University in Bratislava, 84248 Bratislava, Slovakia}
\author{E.~Nacher} %ICC
\affiliation{Instituto de F\'isica Corpuscular, CSIC - Universidad de Valencia, E-46980, Valencia, Spain}
\author{A.~Negret} %ICC
\affiliation{“Horia Hulubei” National Institute for Physics and Nuclear Engineering, RO-077125 Bucharest, Romania}
\author{J.~Ojala}
\affiliation{University of Jyv\"{a}skyl\"{a}, Department of Physics, P.O. Box 35, FI-40014, Jyv\"{a}skyl\"{a}, Finland}
\affiliation{Helsinki Institute of Physics, University of Helsinki, P.O. Box 64, FIN-00014, Helsinki, Finland}
\author{J.~D.~Ovejas}
\affiliation{Instituto de Estructura de la Materia, CSIC, Serrano 113 bis, E-28006 Madrid, Spain}
\author{R.~D.~Page} %ICC
\affiliation{Department of Physics, Oliver Lodge Laboratory, University of Liverpool, Liverpool L69 7ZE, United Kingdom}
\author{P.~Papadakis}
\affiliation{Department of Physics, Oliver Lodge Laboratory, University of Liverpool, Liverpool L69 7ZE, United Kingdom}
\affiliation{STFC Daresbury, Daresbury, Warrington WA4 4AD, United Kingdom}
\author{S.~Pascu} %ICC
\affiliation{“Horia Hulubei” National Institute for Physics and Nuclear Engineering, RO-077125 Bucharest, Romania}
\author{A.~Perea} %ICC
\affiliation{Instituto de Estructura de la Materia, CSIC, Serrano 113 bis, E-28006 Madrid, Spain}
\author{Zs.~Podoly\'ak}
\affiliation{Department of Physics, University of Surrey, Guildford GU2 7XH, United Kingdom}
\author{V.~Pucknell} %ICC
\affiliation{STFC Daresbury, Daresbury, Warrington WA4 4AD, United Kingdom}
\author{E.~Rapisarda} %ICC
\affiliation{ISOLDE, CERN, CH-1211 Geneva 23, Switzerland}
\author{F.~Rotaru} %ICC
\affiliation{“Horia Hulubei” National Institute for Physics and Nuclear Engineering, RO-077125 Bucharest, Romania}
\author{C.~Sotty}%really helpful people
\affiliation{“Horia Hulubei” National Institute for Physics and Nuclear Engineering, RO-077125 Bucharest, Romania}
\author{O.~Tengblad} %ICC
\affiliation{Instituto de Estructura de la Materia, CSIC, Serrano 113 bis, E-28006 Madrid, Spain}
\author{V.~Vedia} %ICC
\affiliation{Grupo de F\'isica Nuclear, Universidad Complutense de Madrid, 28040, Madrid, Spain}
\author{S.~Vi\~nals} %ICC
\affiliation{Instituto de Estructura de la Materia, CSIC, Serrano 113 bis, E-28006 Madrid, Spain}
\author{R.~Wadsworth} %ICC
\affiliation{Department of Physics, University of York, YO10 5DD, York, UK}
\author{N.~Warr} %ICC
\affiliation{Institut f\"{u}r Kernphysik, Universit\"{a}t zu K\"{o}ln, 50937 K\"{o}ln, Germany}
\author{K.~Wrzosek-Lipska}
\affiliation{Heavy Ion Laboratory, University of Warsaw, PL-02-093, Warsaw, Poland}

\collaboration{IDS Collaboration}

\date{\today}

\begin{abstract}
Decay spectroscopy of the long-lived states in $^{186}$Tl has been performed at the ISOLDE Decay Station at ISOLDE, CERN. The $\alpha$ decay from the low-spin $(2^-)$ state in $^{186}$Tl was observed for the first time and a half-life of $3.4^{+0.5}_{-0.4}$ s was determined. Based on the $\alpha$-decay energy, the relative positions of the long-lived states were fixed, with the $(2^-)$ state as the ground state, the $7^{(+)}$ state at 77(56)~keV and the $10^{(-)}$ state at 451(56)~keV. The level scheme of the internal decay of the $^{186}$Tl($10^{(-)}$) state ($T_{1/2} = 3.40(9)$ s), which was known to decay solely through emission of 374~keV $\gamma$-ray transition, was extended and a lower-limit for the $\beta$-decay branching $b_\beta > 5.9(3)\%$ was determined. The extracted retardation factors for the $\gamma$ decay of the $10^{(-)}$ state were compared to the available data in neighboring odd-odd thallium isotopes indicating the importance of the $\pi d_{3/2}$ shell in the isomeric decay and significant structure differences between $^{184}$Tl and $^{186}$Tl.
\end{abstract}

\maketitle

\section{\label{sec:introduction}Introduction}

Neutron-deficient nuclei around the neutron midshell $N=104$ are interesting study cases from the nuclear structure point of view. This region of the nuclear chart is characterized by the occurrence of shape coexistence in atomic nuclei \cite{Heyde2011}, a phenomenon, whereby different shapes coexist within one nucleus at low energy and which is interpreted as arising from proton excitations across the $Z=82$ proton shell closure. These coexisting structures have been observed in laser spectroscopy \cite{Barzakh2013,Barzakh2017,Marsh2018,Sels2019}, $\alpha$- and $\beta$-decay \cite{Andreyev2003,Andreyev2003a,VanBeveren2015,CVanBeveren2016,Rapisarda2017} and Coulomb excitation \cite{Wrzosek-Lipska2019} studies. 

In the case of odd-odd thallium isotopes in this region, proton excitations across $Z=82$ leads to the existence of three long-lived states \cite{VanDuppen1991,CVanBeveren2016}. Previous studies suggest that the main configurations of these states are $[\pi s_{1/2} \otimes \nu p_{3/2}]$ for the $2^-$ states \cite{Andreyev2003,Barzakh2017,VanDuppen1991}, 
a mixture of $[\pi s_{1/2} \otimes \nu i_{13/2}]$ and $[\pi d_{3/2} \otimes \nu i_{13/2}]$ for the $7^+$ states \cite{Andreyev2003,Barzakh2017,VanDuppen1991} and $[\pi h_{9/2} \otimes \nu i_{13/2}]$ for the $10^-$ states ($9^-$ in case of $^{188}$Tl) \cite{Barzakh2017,VanDuppen1991}. Unlike in the neighboring $^{184}$Tl and $^{188}$Tl isotopes, the half-life and the decay of the ($2^-$) state in $^{186}$Tl, whose existence has been suggested from the $\alpha$-decay study of the low-spin isomeric state in $^{190}$Bi \cite{VanDuppen1991}, have not been reported yet. 

Studying the decay pattern of the $10^-$ isomers in Tl isotopes can reveal information on the decay of the intruder based states in this region of the nuclear chart \cite{VanBeveren2015}. However, in the case of $^{186}$Tl, the $10^{(-)}$ state is known to decay only through emission of a 374 keV $\gamma$ ray \cite{Kreiner1981}, while, as observed in $^{184}$Tl, the decay pattern is expected to be more complex, including multiple paths of internal decay, as well as $\alpha$ decay \cite{VanBeveren2015,CVanBeveren2016}.

In this paper we present an extension of the isomeric decay scheme of the $^{186}$Tl($10^{(-)}$) state, the observation of the $\alpha$ decay of the $^{186}$Tl($2^-$) state and the relative positions of the three long-lived states in $^{186}$Tl. The results of the $\beta$-decay study of all three long-lived states will be published elsewhere \cite{StryjczykHgbeta}.

\section{\label{sec:setup}Experimental setup}

The experiment was performed at ISOLDE, CERN as a part of a campaign dedicated to measure the decays of $^{182,184,186}$Tl. A pure beam of $^{186}$Tl was produced through spallation of a thick UC$_x$ target by 1.4 GeV protons, provided by the Proton Synchrotron Booster. The proton pulses (PP) were delivered every 1.2 seconds (or a multiple of this value) and grouped into the CERN proton supercycle structure (SC), whose length varied during the experiment from 20 to 40 PP. The produced thallium atoms effused from the target to a hot cavity, where they were selectively ionized in a two-step ionization process by the Resonance Ionization Laser Ion Source system \cite{Fedosseev2017}. The first step excitation was performed through the 6p $^2$P$_{1/2} \rightarrow$ 6d $^2$D$_{3/2}$ transition at 276.83 nm using a dye laser system. For the second step the output from a Nd:YAG laser at 532 nm was used (details of the laser schemes are given in \cite{Fedosseev2017}). After the ionization, the ions were extracted from the ion source at 30 keV energy and separated with respect to their mass-to-charge ratio by the High Resolution Separator \cite{Catherall2017}. To allow for the implantation of the thallium isotopes, a beam gate was open for 90 ms after each PP. The purified beam was implanted onto an aluminized mylar tape at the center of the ISOLDE Decay Station (IDS) \cite{IDS}. After every SC, the tape was moved in order to remove daughter activities.

\begin{figure}
\includegraphics[width=\columnwidth]{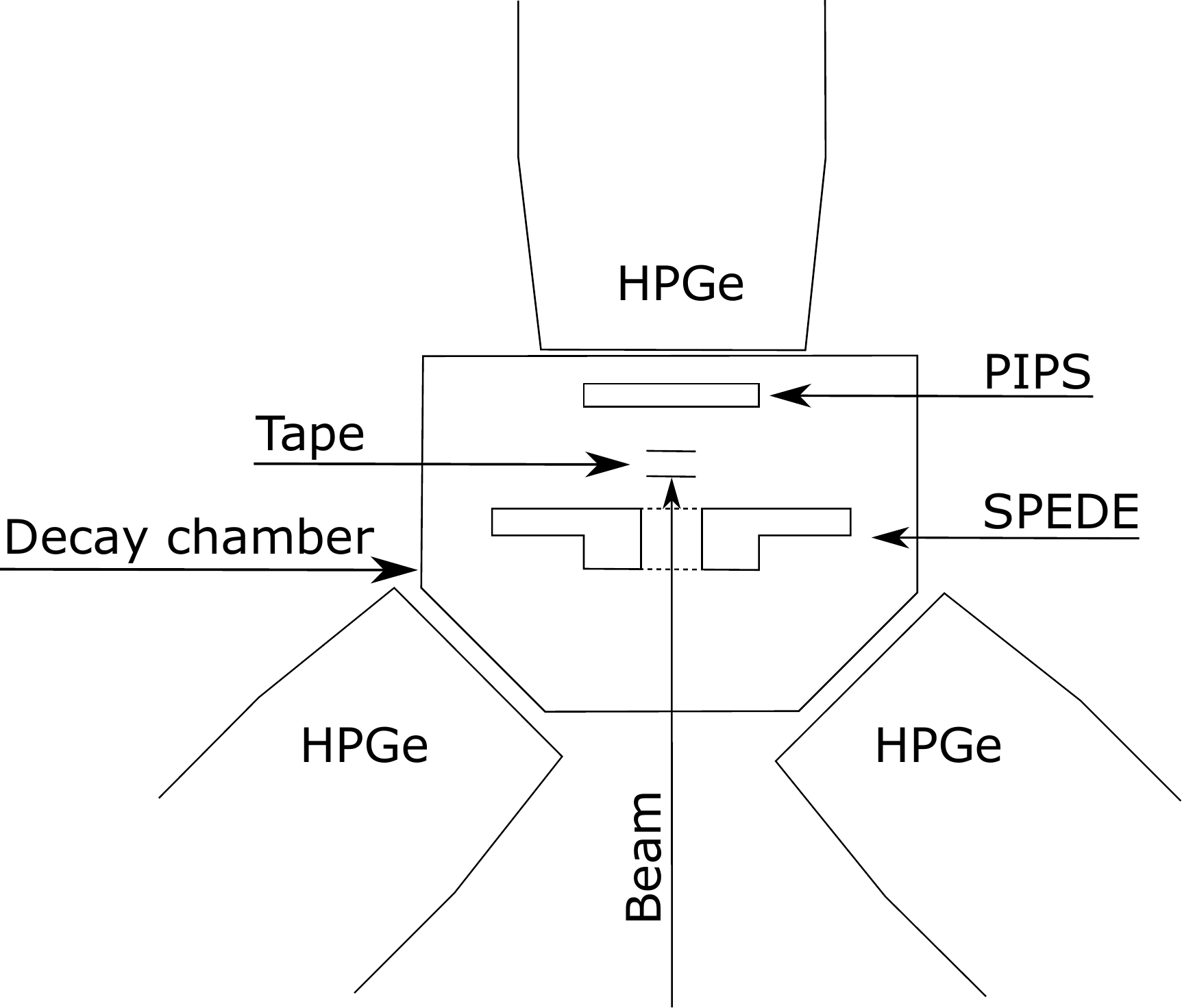}
\caption{\label{fig:idschamber}A scheme of the detection system.}
\end{figure}

The SPEDE spectrometer was installed in the IDS decay chamber for the detection of conversion electrons and $\alpha$ particles \cite{Papadakis2018}. It consists of 24-fold segmented, 1-mm thick annular silicon detector, which was cooled by circulating ethanol at about $-20^{\mathrm{o}}\mathrm{C}$. It was situated at 15 mm distance in the backward direction of the beam, in front of the tape (see Fig. \ref{fig:idschamber}) and it covered about 14\% of the solid angle. Behind the tape, a 0.5-mm thick 900 mm$^2$ PIPS silicon detector for the detection of $\beta$ particles was placed. Outside the IDS chamber, there were five High-Purity Germanium Clover detectors (HPGe) used to detect the $\gamma$ radiation. The $\gamma$ energy and efficiency calibrations were performed by using an encapsulated $^{152}$Eu source and a $^{138}$Cs sample produced on-line and implanted onto the tape. At 1408~keV the energy resolution was 2.7~keV and the absolute $\gamma$ efficiency was 1.95(6)\%. The SPEDE spectrometer energy calibration was performed using strong transitions with known energies from the decays of $^{182,184,186}$Tl and $^{138}$Cs for electrons and by using known $\alpha$ decays of $^{184}$Tl and $^{184}$Hg for $\alpha$ particles. The electron energy resolution was 6.3~keV at 288~keV and the $\alpha$ energy resolution was about 190~keV at 6~MeV. All signals were collected in triggerless mode by using the Nutaq digital acquisition system \cite{Nutaq} with 100 MHz sampling frequency.

\section{\label{sec:results}Results}

\subsection{\label{sec:state2}$\alpha$ decay of the ($2^-$) state}

\begin{figure}
\includegraphics[width=\columnwidth]{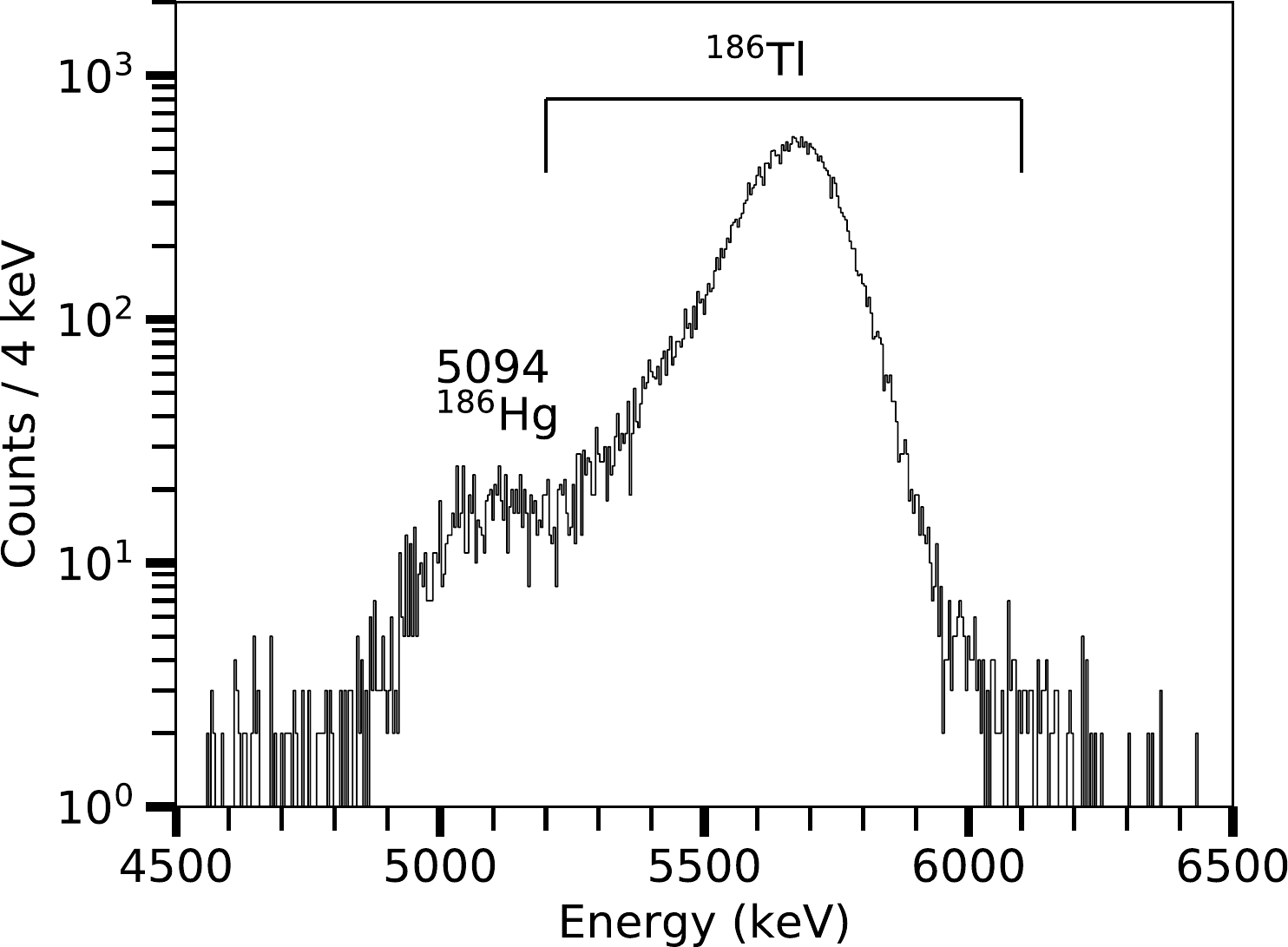}
\caption{\label{fig:alphasingles}The single-$\alpha$ energy spectrum registered with the SPEDE spectrometer.}
\end{figure}

Figure \ref{fig:alphasingles} shows the $\alpha$-decay energy spectrum registered by the SPEDE spectrometer during the experiment. Due to a limited energy resolution, it was not possible to resolve the fine structure $\alpha$ decays. Based on the: (i) systematic trend of the $\alpha$-particle energies (see Figs. 1 and 9 in Ref. \cite{CVanBeveren2016}), (ii) the previous experimental measurements of the $\alpha$ decay of $^{186}$Tl and $^{186}$Hg \cite{Ijaz1977,Hansen1970}, and (iii) the behavior of the peak intensity as a function of time, the peak around 5.7 MeV was associated with the $\alpha$ decay of $^{186}$Tl, while the peak at around 5.1 MeV stems from a weak ground-state to ground-state $\alpha$-decay of $^{186}$Hg to $^{182}$Pt (branching ratio $b_\alpha = 0.016(5) \%$, $E_\alpha = 5094(15)$ keV \cite{Hansen1970}).
%  \cite{Hansen1970}) with a known energy of $E_\alpha = 5094(15)$ keV \cite{Hansen1970}.

\begin{figure}
\includegraphics[width=\columnwidth]{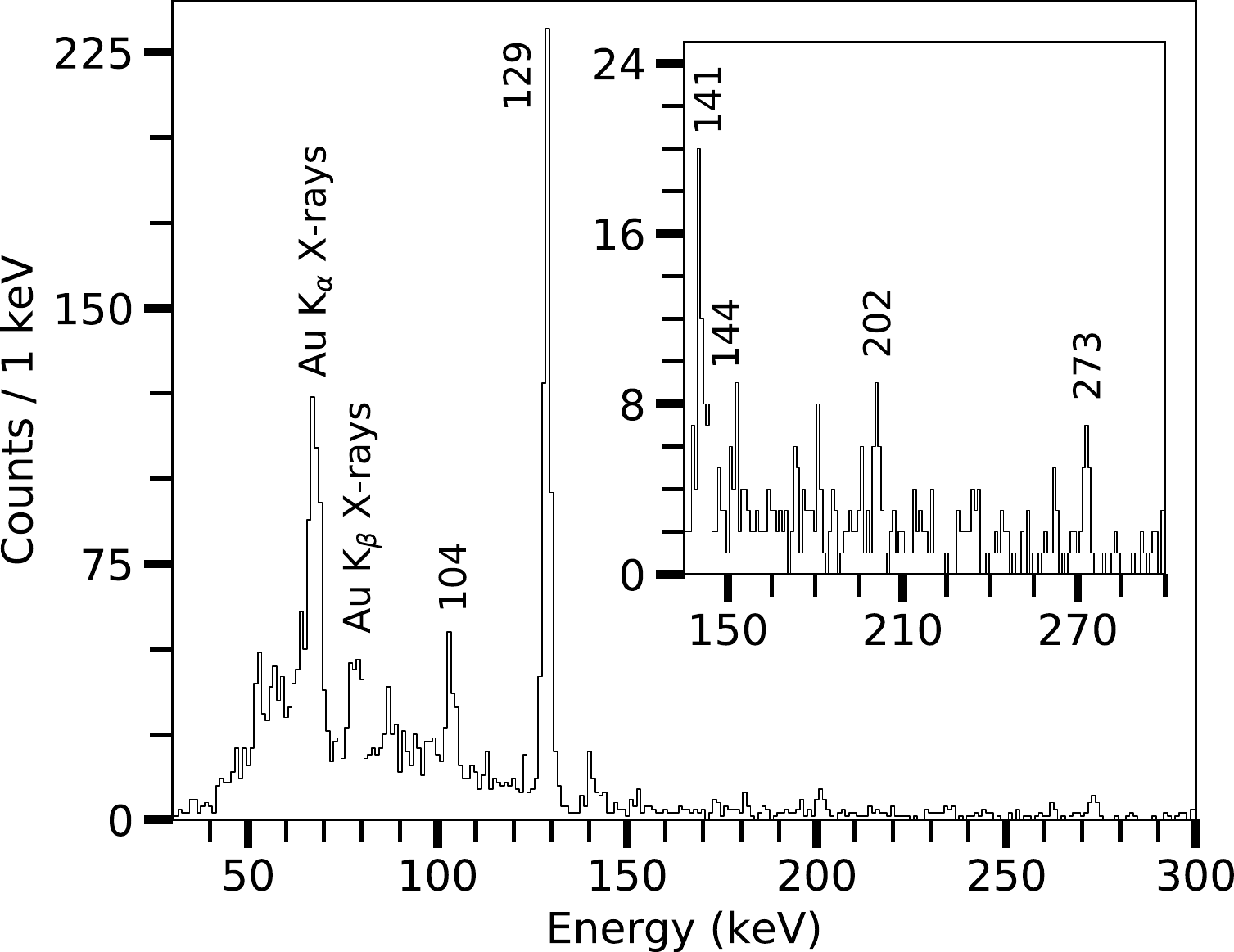}
\caption{\label{fig:ag}The $\gamma$-ray energy spectrum with an energy gate on the 4550 to 6500 keV $\alpha$ particles and a coincidence time gate $70 \mathrm{~ns} \leq \Delta T(\gamma \mathrm{-} \alpha) \leq 300 \mathrm{~ns}$. An expanded view of the spectrum between 135 keV and 300 keV is shown in the inset. Peaks are labeled according to corresponding transition energy in keV.}
\end{figure}

\begin{figure}
\includegraphics[width=\columnwidth]{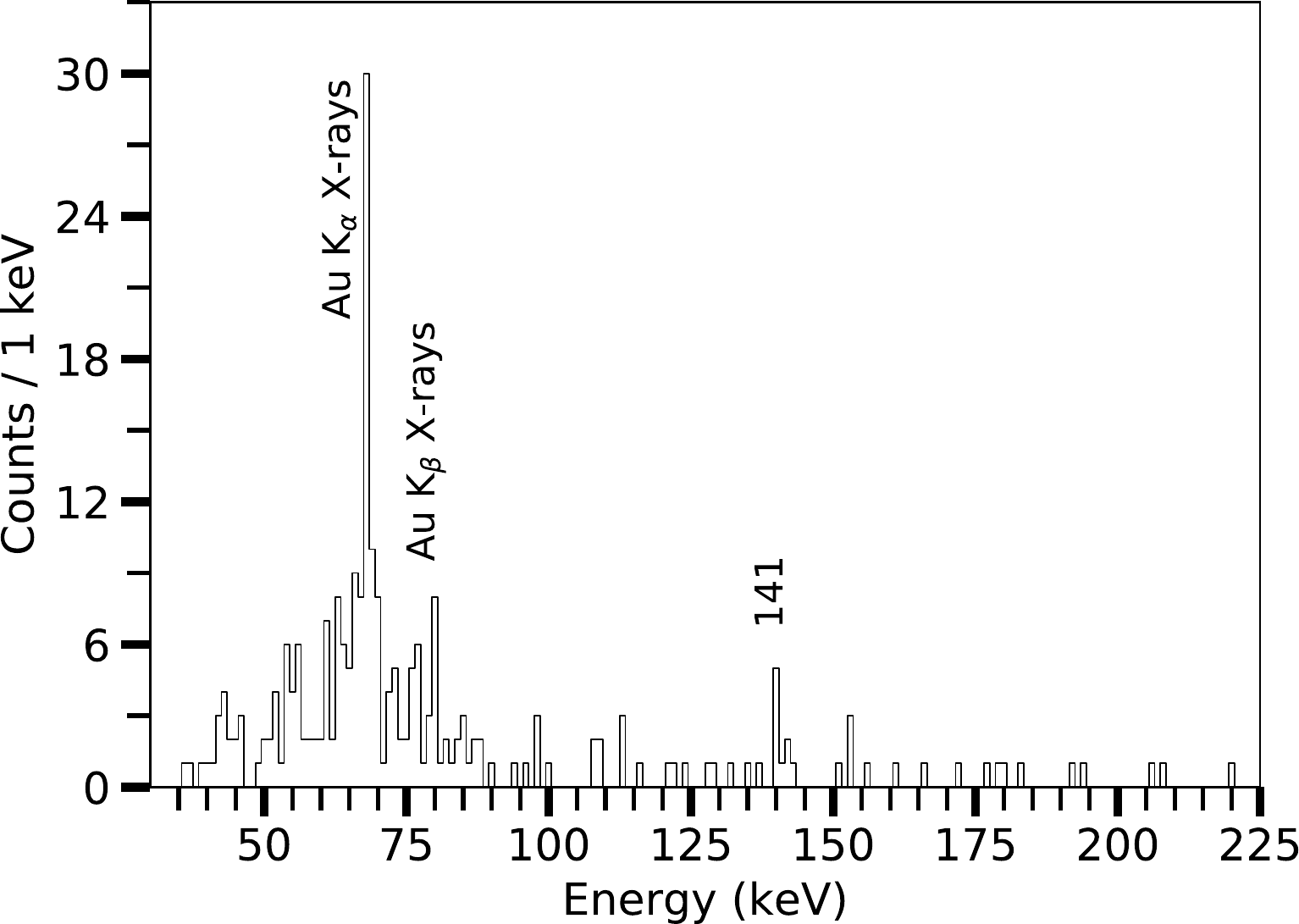}
\caption{\label{fig:agg}The $\gamma$-ray energy spectrum with an energy gate on the 4550 to 6500 keV $\alpha$ particles and 129 keV $\gamma$ ray. Gold K$_\alpha$ and K$_\beta$ X-rays as well as a peak at 141 keV are present.}
\end{figure}

\begin{figure}
\includegraphics[width=\columnwidth]{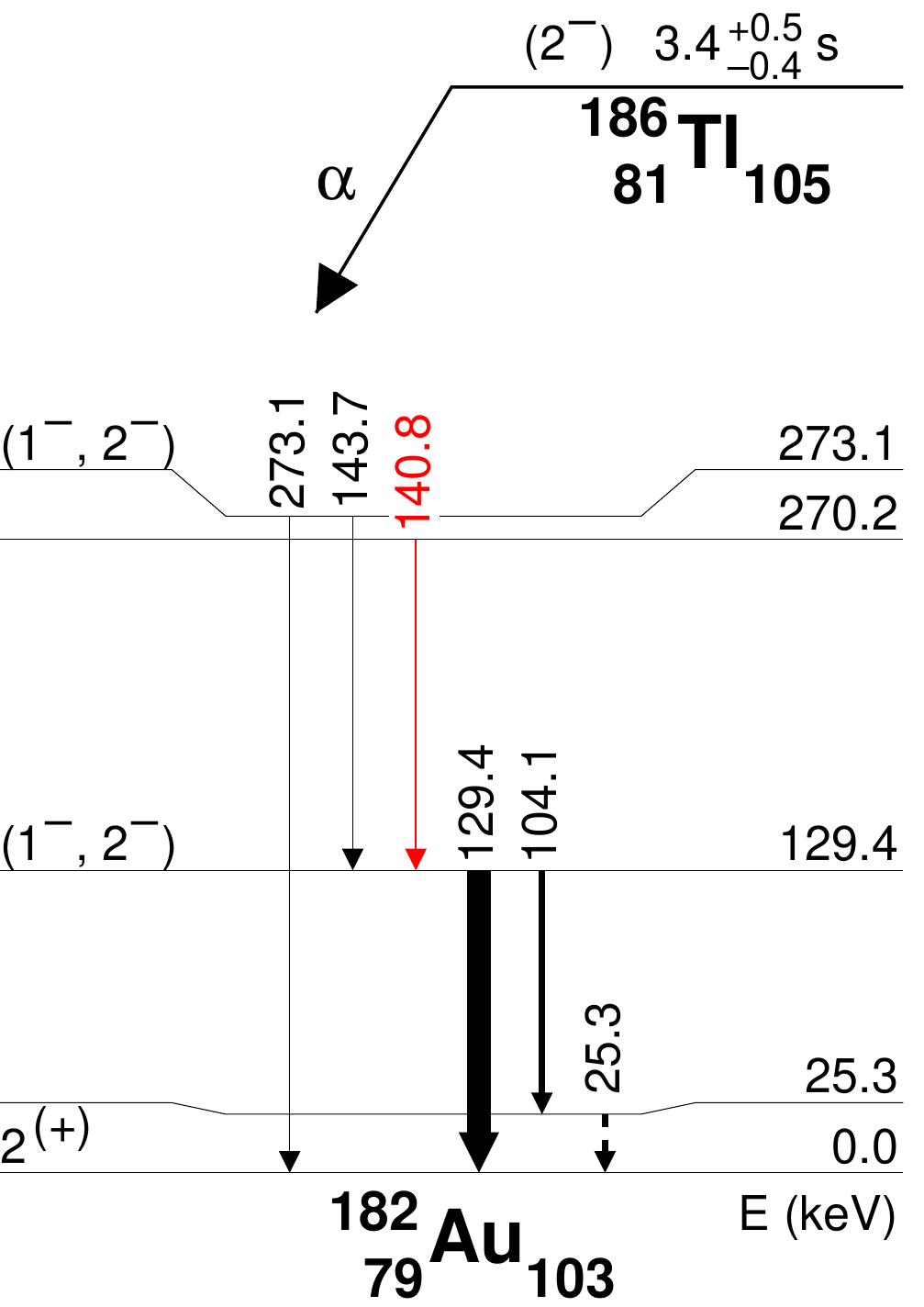}
\caption{\label{fig:scheme182Au}The $\alpha$-decay scheme obtained in the present work. The 25 keV transition has not been observed, however, it is known from the $\beta$-decay studies \cite{Ibrahim2001}. The spin-parity of the $^{182}$Au ground state is taken from Ref. \cite{Harding2020} while the spin assignments of the excited states and the placement of the $\gamma$-ray transitions are taken from Ref. \cite{Ibrahim2001}, with an exception of the newly observed 141 keV transition (plotted in red). The spin-parity of $^{186}$Tl is taken from Ref. \cite{VanDuppen1991} and the half-life comes from our analysis.}
\end{figure}

In total, six $\gamma$-ray transitions have been observed in coincidence with the $\alpha$ particles associated with the decay of $^{186}$Tl (see Table \ref{tab:gammas182Au} and Fig. \ref{fig:ag}). Four of them, 104 keV, 129 keV, 144 keV and 273 keV transitions, were previously observed and placed in the level scheme in the $^{182}$Hg to $^{182}$Au $\beta$-decay study \cite{Ibrahim2001}. Based on the $\alpha$-$\gamma$-$\gamma$ coincidences (see Fig. \ref{fig:agg}), the newly identified 141 keV transition was placed on top of the 129 keV level, while the 202 keV transition remains unplaced. The decay scheme is presented in Fig. \ref{fig:scheme182Au}.

%From the $^{182}$Hg to $^{182}$Au $\beta$-decay study \cite{Ibrahim2001}, four of them, 104 keV, 129 keV, 144 keV and 273 keV, are already known and placed in the decay scheme of $^{182}$Au.

Two levels at 129 keV and 273 keV known from the $\beta$-decay study of $^{182}$Hg($0^+$)  are both suggested to have spin 1 or 2 and a negative parity \cite{Ibrahim2001}. The feeding of these low-spin levels in the $\alpha$ decay of $^{186}$Tl suggests a low spin for the $\alpha$-decaying state, despite the fact that this state has a similar half-life to the 10$^{(-)}$ state (see Sec. \ref{sec:halflives}). The existence of such level, with spin-parity ($2^-$), has been proposed from the $\alpha$-decay studies of $^{190}$Bi \cite{VanDuppen1991}. Therefore, we suggest that the observed $\gamma$-ray transitions placed in the decay scheme in Fig. \ref{fig:scheme182Au} follow the $\alpha$ decay of the $^{186}$Tl($2^-$) state.

\begin{table}
\caption{\label{tab:gammas182Au}
The relative intensities of the $\gamma$-ray transitions assigned to the decay of $^{186}$Tl to $^{182}$Au, normalized to the intensity of the 129 keV transition. The 202 keV transition remains unplaced on the decay scheme.}
\begin{ruledtabular}
\begin{tabular}{cccc}
$E_{\gamma}$			& $I_\gamma^{\mathrm{rel}}$		& $E_{\mathrm{level}}^{\mathrm{initial}}$ 	& $E_{\mathrm{level}}^{\mathrm{final}}$  	\\
(keV) 					& 						& (keV) 		& (keV) 		\\\hline
104.1(2) 				& 24$^{+5}_{-4}$ 	& 129.4  	& 25.3		\\
129.4(1) 				& 100 					& 129.4  	& 0.0		\\
140.8(3) 				& 6.1$^{+1.7}_{-1.5}$	& 270.2  	& 129.4		\\
143.7$^{+0.7}_{-0.8}$ 	& 2.1$^{+1.2}_{-1.0}$	& 273.1  	& 129.4		\\
201.5$^{+0.4}_{-0.5}$ 	& 2.4$^{+1.0}_{-0.8}$	& -  		& -			\\
273.1$^{+0.5}_{-0.7}$ 	& 3.0$^{+1.0}_{-0.8}$	& 273.1 		& 0.0		\\
\end{tabular}
\end{ruledtabular}
\end{table}

%I do not think this will be clear to people not involved. I would first explicitly state something like "Furthermore, the 104 keV transition was not used for gating, because ...". And only then "To account for feeding by this transition, ..."

To check the possible $\alpha$ decays of the other long-lived states in $^{186}$Tl, the number of counts in the $\alpha$-decay spectra gated on the 129 keV and 273 keV $\gamma$-ray transitions were compared to the number of counts in the single-$\alpha$ energy spectrum. To remove the influence of the 5094 keV $\alpha$ particles from the $^{186}$Hg decay, the comparison range was set between 5.4 and 6.5 MeV. Both $\gamma$-gated spectra were corrected by the detection efficiency and by the total conversion coefficient calculated using BrIcc \cite{Kibedi2008}. The 202 keV transition was not included because it is unplaced in the decay scheme while the 141 keV and 144 keV transitions were not included to avoid double counting of the $\alpha$ particles since they feed the 129 keV level. Furthermore, the 104 keV transition was not used for gating, because it may also originate from the decay of the ($7^+$) state in $^{182}$Au \cite{Zhang2002}. To account for this feeding, the $\alpha$ energy spectrum gated on the 129 keV transition was corrected by the total intensity of the 104 keV transition, 32(3)\%, from the $^{182}$Hg $\beta$-decay studies \cite{Ibrahim2001}. From this comparison, the $\gamma$-gated-$\alpha$ counts can reproduce 105(7)\% of the total number of $\alpha$ counts in the single-$\alpha$ energy spectrum suggesting that the vast majority of the registered $\alpha$ decays originates from the ($2^-$) state (see Sec. \ref{sec:state2discussion}). 

%(see Sec. \ref{sec:state2discussion} for details) 

%To check the possible $\alpha$ decays of the other long-lived states in $^{186}$Tl, the number of counts in the $\gamma$-gated-$\alpha$ spectrum corrected by the $\gamma$-detection efficiency was compared to the number of counts in the single-$\alpha$ spectrum. To remove the influence of the 5094 keV $\alpha$ particles originating from the decay of $^{186}$Hg, the comparison range was set between 5.4 and 6.5 MeV. To avoid double counting of the $\alpha$ particles, the 141 keV and 144 keV transitions were not included since they feed the 129 keV level. The counts gated by the 104 keV transition could not be included, since the 104 keV $\gamma$ rays may also originate from the decay of the ($7^+$) state in $^{182}$Au \cite{Zhang2002} (see Sec. \ref{sec:state2discussion} for details). To account for the feeding from the 104 keV transition, the number of counts gated on the 129 keV transition was increased by the intensity of the 104 keV transition, 23(2)\%, obtained from the $\beta$-decay studies of $^{182}$Hg \cite{Ibrahim2001}. The 202 keV transition was not included because it is unplaced in the decay scheme and may originate from the decay of the high-spin isomers in $^{186}$Tl. From this comparison, the $\gamma$-gated-$\alpha$ counts can reproduce 85(4)\% of the total number of $\alpha$ counts in the single spectrum suggesting that the majority of the registered $\alpha$ decays originates from the ($2^-$) state (see Sec. \ref{sec:state2discussion}). 

\begin{figure}
\includegraphics[width=\columnwidth]{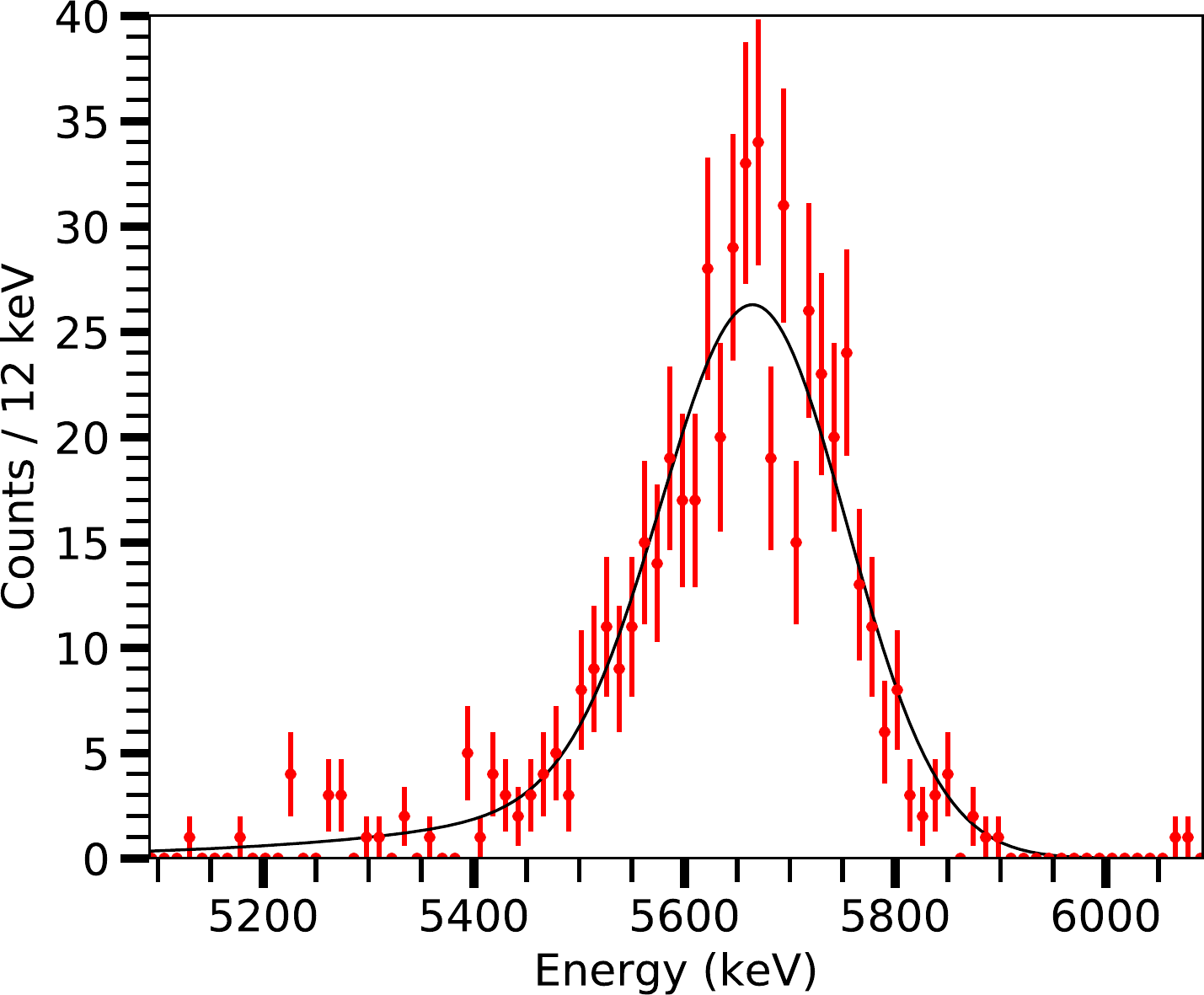}
\caption{\label{fig:ealpha} The $\alpha$-decay energy spectrum gated on the 129 keV transition, plotted together with a fit to the experimental data ($E_\alpha = 5670(51)$ keV).}
\end{figure}

Based on the measured $\alpha$-particle energies with a gate on the 129 keV $\gamma$-ray transition, the energy of the $\alpha$ decay feeding the 129 keV level was determined to be $E_\alpha = 5670(51)$ keV (see Fig. \ref{fig:ealpha}) and it corresponds to the $Q_{\alpha,tot} = Q_{\alpha} + E_\gamma =  5924(52)$ keV. Due to limited statistics, it was not possible to fit the $\alpha$ energies feeding other states. The extracted $Q_{\alpha,tot}$ and the atomic masses of $^{4}$He and $^{182}$Au \cite{Wang2017,Pfeiffer2014} allowed us to calculate the atomic mass of the $^{186}$Tl($2^-$) state 185978582(60) $\mu$u and to compare it with the atomic mass of the $^{186}$Tl($7^{(+)}$) state (185978664.2(67) $\mu$u \cite{Weber2008,Pfeiffer2014}). Our analysis indicates that the ($2^-$) state is the ground state and the $7^{(+)}$ level has an excitation energy of 77(56) keV. Based on our result and the $\alpha$-particle energies \cite{VanDuppen1991,Andreyev2003}, it was also possible to fix the relative positions of the $\alpha$-decaying states in $^{190}$Bi with the ($10^-$) level being 182(57) keV above the ($3^+$) ground state.

%This result also allowed us to fix the relative positions of the $\alpha$-decaying states in $^{190}$Bi with the ($10^-$) level being 182(57) keV above the ($3^+$) ground state.

% ($3^+$) level being the ground state and the ($10^-$) level having an excitation energy of 182(57) keV.

%, which was measured at ISOLTRAP, ISOLDE \cite{Weber2008,Pfeiffer2014}. 185978664,2	6,7
% (181969618(21) $\mu$u
%(4002603.25413(6) $\mu$u \cite{Wang2017})

%The influence of the transitions feeding the 129 keV level on the fit of the $\alpha$ energy was checked and it is negligible as these transitions are much weaker compared to the 129 keV transition. 

%. The extracted value was used to calculate the $Q_{\alpha,tot} = Q_{\alpha} + E_\gamma =  5924(52)$ keV. 

\subsection{\label{sec:state10}Decay of the $10^{(-)}$ state}

\begin{figure*}
\includegraphics[width=\textwidth]{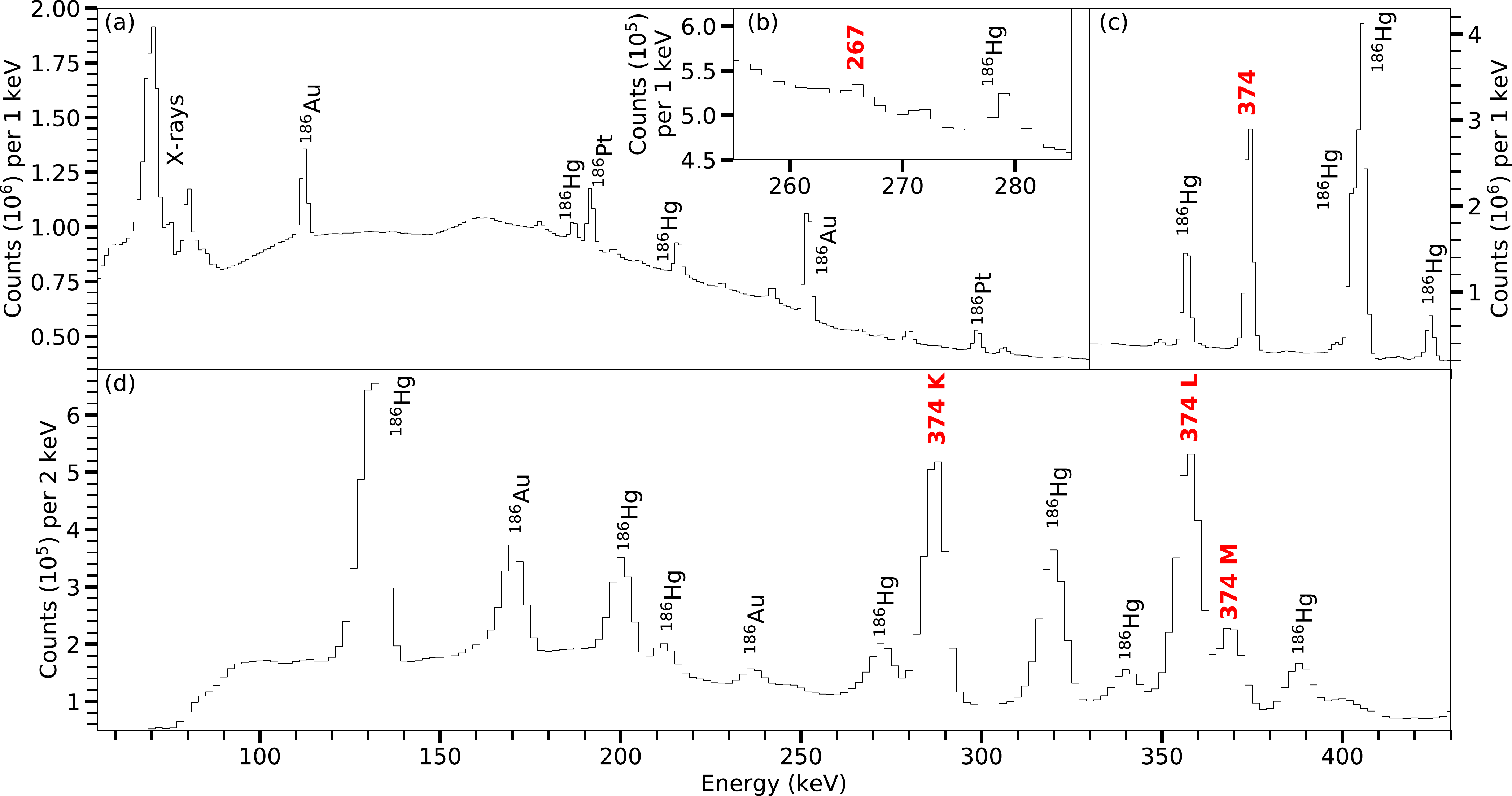}
\caption{\label{fig:singles}Portions of the single-$\gamma$ (panels a, b and c) and single-electron (panel d) energy spectra. The transitions from the decay of the $^{186}$Tl($10^-$) isomer are labeled in red. Other strong peaks are labeled with the symbol of the nucleus they originate from. Note different scales on the y-axes of each panel.}
\end{figure*}

The single $\gamma$-ray energy spectrum (Fig. \ref{fig:singles}) shows the 374 keV transition previously assigned to the $10^{(-)} \rightarrow 7^{(+)}$ decay \cite{Kreiner1981}. By using $\gamma$-$\gamma$ coincidences, another decay cascade deexciting through an 18 keV transition was established. Furthermore, a $\beta$-decay channel of the $10^{(-)}$ state was identified.

%of the $10^{(-)}$ state
%dodać info wprost o istnieniu 18 keV i beta decay, które jest omówione w kolejnych paragrafach. 

\begin{figure}
\includegraphics[width=\columnwidth]{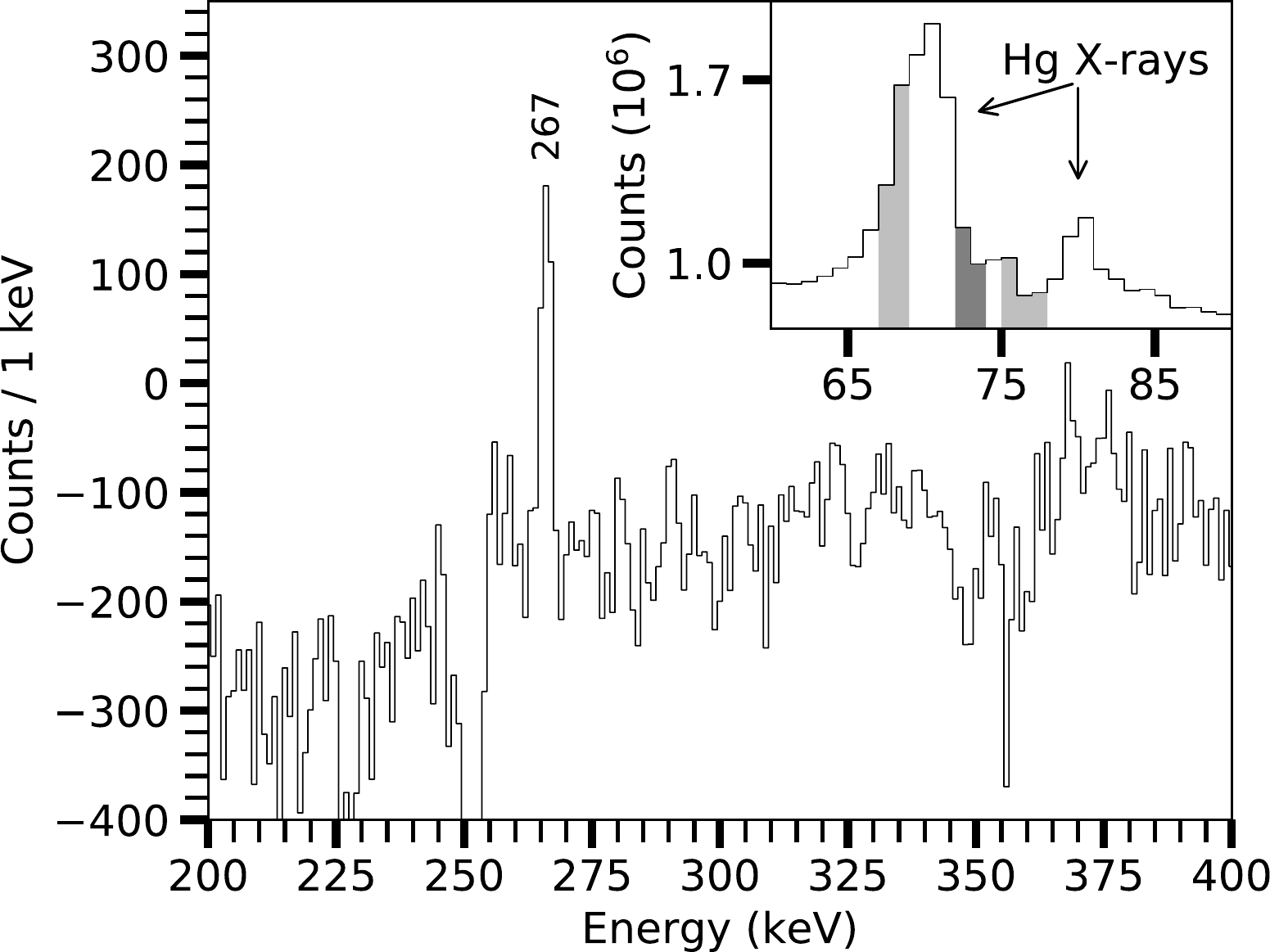}
\caption{\label{fig:ggxray}The background-subtracted $\gamma$-$\gamma$ spectrum gated on the thallium X-rays. A negative background is related to much higher intensity of the Hg K$_\alpha$ X-rays. Two dips at 251 keV and 356 keV are related to the subtraction of the strong $\gamma$-rays from the decay of $^{186}$Hg to $^{186}$Au and $^{186}$Tl to $^{186}$Hg, respectively. In the inset, a portion of the single-$\gamma$ energy spectrum with the gate region (dark gray) and the background region (light gray) used to create the $\gamma$-ray energy spectrum shown, is plotted.}
\end{figure}

\begin{figure}
\includegraphics[width=\columnwidth]{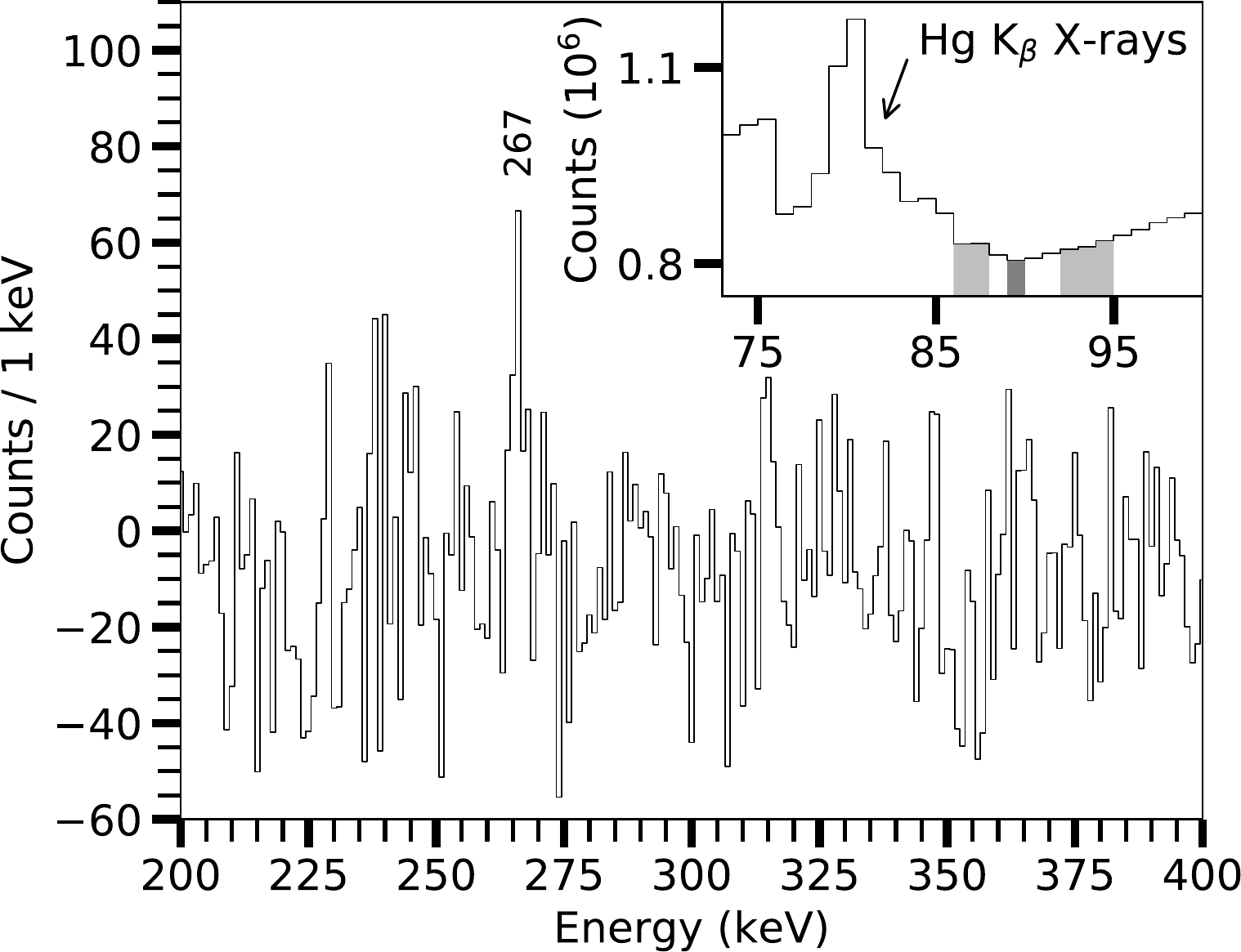}
\caption{\label{fig:gg90}The background-subtracted $\gamma$-$\gamma$ spectrum gated on the $\gamma$ rays between 89 and 90 keV. In the inset, a portion of the single-$\gamma$ energy spectrum with the gate region (dark gray) and the background region (light gray) used to create the $\gamma$-ray energy spectrum shown, is plotted.}
\end{figure}

\begin{figure}
\includegraphics[width=\columnwidth]{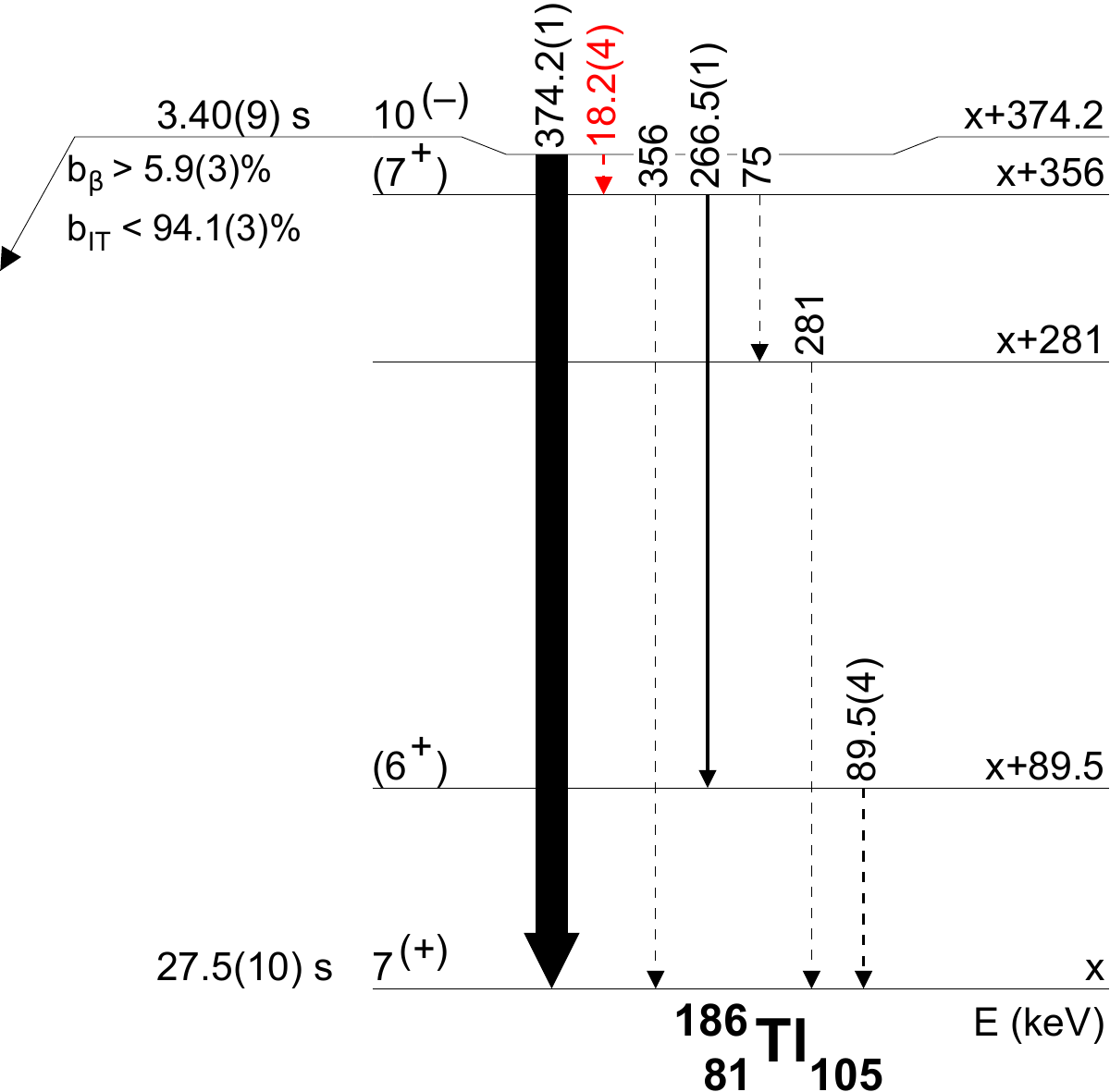}
\caption{\label{fig:scheme186Tl}The decay scheme of the isomeric 10$^{(-)}$ state in $^{186}$Tl obtained from our analysis. The transitions marked with the dashed lines have not been observed, but they are known from the $\alpha$-decay studies \cite{VanDuppen1991,Andreyev2003}. The existence of the 18 keV transition (indicated by the dashed red line) is suggested by the observation of the 267 keV transition in our study. The spin assignments come from Refs. \cite{VanDuppen1991,Andreyev2003,Barzakh2013} except for the $x + 356$ keV state, see text for details. The half-life of the 10$^{(-)}$ state comes from our analysis, while the half-life of the 7$^{(+)}$ state is taken from Ref. \cite{Baglin2003}. From our analysis, $x=77(56)$ keV.}
\end{figure}

The 267 keV $\gamma$-ray transition has been observed in coincidence with thallium K$_\alpha$ X-rays (Fig. \ref{fig:ggxray}), as well as in coincidence with a $\gamma$-ray energy gate set between 89 and 90 keV (Fig. \ref{fig:gg90}). These results are consistent with the $\alpha$-decay study of $^{190}$Bi, where the 267 keV $\gamma$-ray transition was proposed to deexcite the $x+356$ keV level to the $x+89.5(4)$ keV state \cite{VanDuppen1991,Andreyev2003}. These observations also imply that the $x+374$ keV level deexcites through an 18 keV transition to the $x+356$ keV state. In the present work, the other $\gamma$-rays deexciting the $x+356$ keV level \cite{Andreyev2003} were not observed; the 75 keV transition overlaps with the background lead K$_\alpha$ X-rays, while the 281 keV and the 356 keV transitions overlap with strong $\gamma$ rays in $^{186}$Hg. The proposed decay scheme is presented in Fig. \ref{fig:scheme186Tl}. 

%In our analysis, we were not able to observe other $\gamma$-ray transitions observed in the $\alpha$-decay study deexciting the $x+356$ keV level 

The $\gamma$-ray intensities of the 267 keV and 374 keV transitions were extracted from the single $\gamma$-ray energy spectrum (see Fig. \ref{fig:singles} and Table \ref{tab:gammas186Tl}). To estimate the total intensity of the 18 keV transition, the spectrum shown in Fig. 6c of Ref. \cite{Andreyev2003} was analyzed. The peaks at 267 keV, 281 keV and 356 keV contained $\approx$60 counts, $\approx$20 counts and $\approx$54 counts, respectively. The detection efficiencies for the 281 keV and the 356 keV transitions were 98\% and 86\% relative to the efficiency for the 267 keV transition, respectively, following the calibration provided in Ref. \cite{BorisThesis}. To estimate the contribution from the internal conversion, all three transitions were considered to be pure $M1$ or pure $E2$ which resulted in the total feeding of the $x+356$ keV state being equal to 1.8(2)\% and 1.4(2)\%, respectively. Since the true transition multipolarities are not known, 1.6(4)\% was adopted as the total intensity of the 18 keV transition.

\begin{table}
\caption{\label{tab:gammas186Tl}
The relative intensities of the $\gamma$-ray transitions assigned to the internal decay of the $10^{(-)}$ state in $^{186}$Tl.}
\begin{ruledtabular}
\begin{tabular}{ccccc}
$E_{\gamma}$ 				& $I_\gamma^{\mathrm{rel}}$							& Multipolarity	& $\alpha_{\mathrm{tot}}$								& $I_{\mathrm{tot}}^{\mathrm{rel}}$\\
(keV)							& 													& 					& 												& \\\hline	
18.2(4) 						& $4.8(14) \times 10^{-7}$\footnotemark[1]	& $E3$				& $3.3(5) \times 10^6$ \footnotemark[2] 	& 1.6(4)\footnotemark[1]	\\
266.5(1) 						& 0.55(3)											& $M1+E2$		& 0.35(20)\footnotemark[3] 				& 0.74(12) \\
374.2(1)						& 100												& $E3$				& 0.249(4)\footnotemark[2]					& 124.9(4) \\
\end{tabular}
\end{ruledtabular}
\footnotetext[1]{Estimated based on the data presented in \cite{Andreyev2003}, see text for details.}
\footnotetext[2]{Calculated using BrIcc \cite{Kibedi2008} assuming given multipolarity.}
\footnotetext[3]{Average value of conversion coefficient for pure $M1$ and pure $E2$ transition, calculated using BrIcc \cite{Kibedi2008}.}
\end{table}

To determine the possible $\beta$-decay branch from the 10$^{(-)}$ state in $^{186}$Tl, the feeding to the known high-spin states ($\geq 9$) in $^{186}$Hg was analyzed. It was assumed that both, direct and indirect, feeding of these states originates only from the decay of the $^{186}$Tl 10$^{(-)}$ state. In total, six high-spin states in $^{186}$Hg have been observed in our study: the $10^+$ state at 2078.1 keV, the ($9^-$) at 2427.6 keV, the ($9$) at 2573.8 keV, the $12^+$ at 2620.1 keV, the ($10^+$) at 2636.4 keV and the $10^+$ at 2833.6 keV. By comparing the feeding to these states with the isomeric decay, a $\beta$ branching equal to $5.9(3)\%$ was extracted. It should be noted that the presented method allows us to estimate only the lower limit for the $\beta$ branching since we observed decays of several states with the unknown spins \cite{StryjczykHgbeta}, which can be also fed through the decay of the 10$^{(-)}$ state.

%$b_\beta = 5.9(3)\%$
\subsection{\label{sec:halflives}Half-lives of the ($2^-$) and $10^{(-)}$ states}

\begin{figure}
\includegraphics[width=\columnwidth]{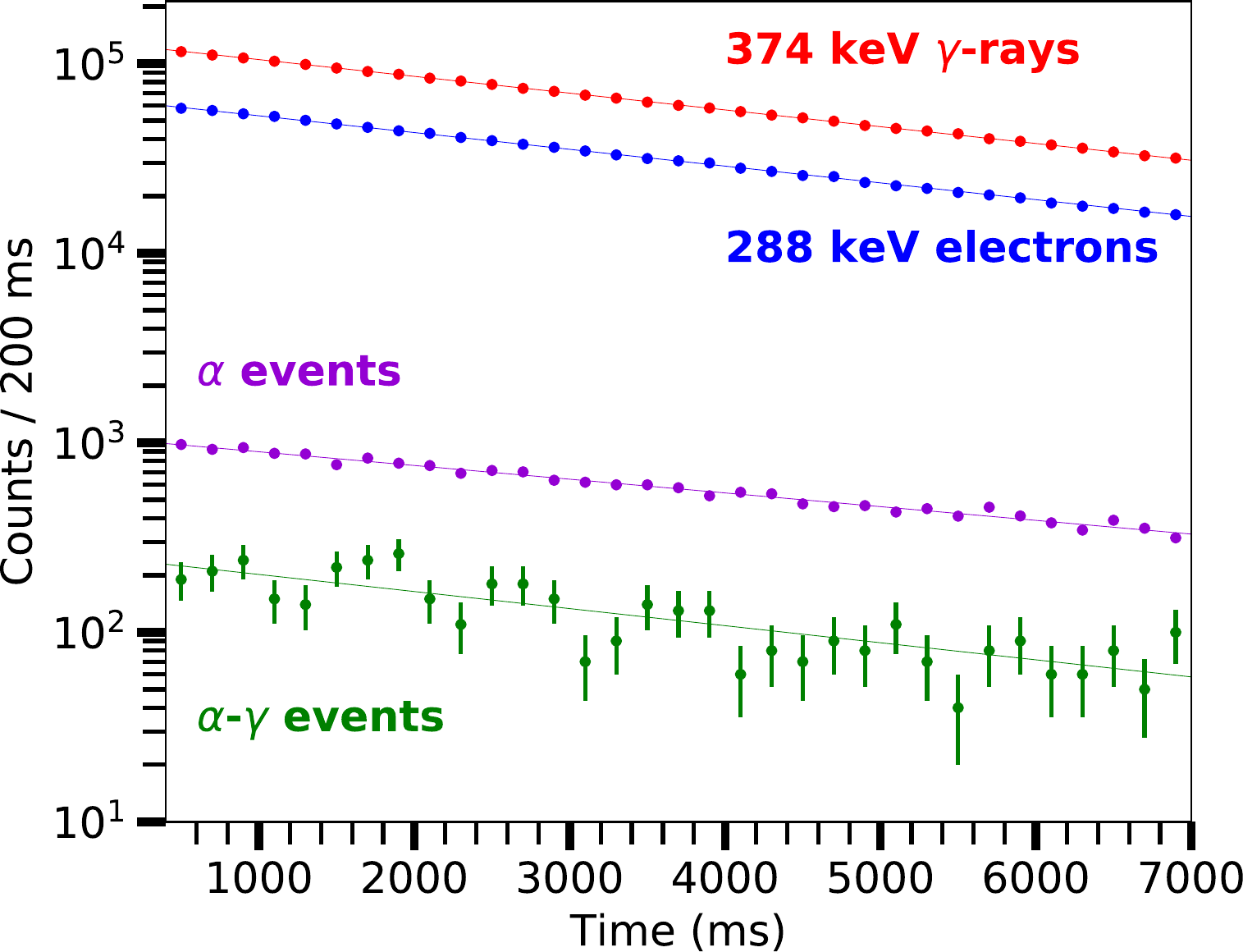}
\caption{\label{fig:halflives} (Color online) The decay curves of the $^{186}$Tl($10^{(-)}$) state ($T_{1/2}(10^{(-)}) = 3.40(9)$~s) fitted to the 374 keV $\gamma$ ray (red) and to the 288 keV electrons (K conversion electrons of the 374 keV transition; blue), as well as decay curves of the $^{186}$Tl($2^-$) state ($T_{1/2}(2^-) = 3.4^{+0.5}_{-0.4}$~s) fitted to the $\alpha$-$\gamma$ events (green, multiplied by a factor 10 for the presentation), and of the $\alpha$ particles in between 5.4 and 6.5 MeV (purple, $T_{1/2} = 4.16(10)$~s).}
\end{figure}

%I do not think that this statement makes sense at the closer look. Maybe like this: "... to the number of  \gamma rays and the K conversion electrons stemming from the 374 keV transition, plotted as a function of time." Also later on, consider if the expressions should not be modified. On one hand it would be more cumbersome to always say "fitted to the number of ...", on the other hand I am not sure if it is correct as it is now.

The half-life of the $10^{(-)}$ state was obtained by simultaneously fitting an exponential function to the number of the $\gamma$ rays and the K conversion electrons stemming from the 374 keV transition (Fig. \ref{fig:singles}) plotted as a function of time (Fig. \ref{fig:halflives}). The fitting range was set from 400 to 7000 ms after the PP and the likelihood function was built assuming that all the points are following a Poisson distribution. The results of the fit are presented in Fig. \ref{fig:halflives} (red and blue curves). The obtained value of $T_{1/2}(10^{(-)}) = 3.40(9)$~s is in agreement with $4.5^{+1.0}_{-1.5}$~s reported in Ref. \cite{Hamilton1975}, $4.5(13)$~s reported in Ref. \cite{Cole1977} and $3(1)$ s reported in Ref.  \cite{Beraud1977}, however, it is more than $2\sigma$ away from the 2.9(2)~s reported in Ref. \cite{Kreiner1981}.

The half-life of the ($2^-$) state was obtained by fitting an exponential function to the $\alpha$-$\gamma$ events plotted as a function of time. The energy gates were set on all the $\gamma$ rays placed in decay scheme of the $^{186}$Tl($2^-$) state, with the exception of the 104 keV transition, as it might also originate from the decay of the $7^{(+)}$ state (see Sec. \ref{sec:state2discussion}). The fitting range and the method are the same as for the half-life of the $10^{(-)}$ state. The results of the fit are presented in Fig. \ref{fig:halflives} (green curve). The extracted half-life is equal to $3.4^{+0.5}_{-0.4}$~s. This value was compared to the results obtained from the fitting of the exponential function to the time distribution of the $\alpha$ particles with energies between 5.4 and 6.5 MeV (Fig. \ref{fig:halflives}, purple curve). The extracted half-life of $4.16(10)$~s is larger, but in agreement within 2$\sigma$. The discrepancy might be explained by a small admixture of the $\alpha$ particles from the long-lived $^{186}$Tl($7^{(+)}$) state ($T_{1/2} = 27.5(10)$~s \cite{Baglin2003}), thus we adopted $T_{1/2} = 3.4^{+0.5}_{-0.4}$~s as the half-life of the ($2^-$) state. However, we note that the expected admixture should be small, which was presented by comparing the number of single-$\alpha$ and $\alpha$-$\gamma$ events (see Sec. \ref{sec:state2discussion}).

%a limited statistics as, well as

%The value obtained from the $\alpha$-$\gamma$ events was compared to the results obtained from the fitting of the exponential function to the time distribution of the $\alpha$ particles with energies between 5.4 and 6.5 MeV (Fig. \ref{fig:halflives}, purple curve). The extracted half-life of $4.16(10)$~s is larger but in a 2$\sigma$ agreement with the val, which might suggest an influence of a long-lived activity, namely an $\alpha$ decay from the $7^{(+)}$ state ($T_{1/2}(7^+) = 27.5(10)$~s \cite{Baglin2003}). This hypothesis is further supported by the fact that the half-life obtained by fitting two exponential functions, with one half-life left as a free parameter and the second one given as a prior equal to 27.5(10) s, is equal to $3.7^{+0.3}_{-0.4}$~s, which is in a good agreement with $T_{1/2}(2^-) = 3.4^{+0.5}_{-0.4}$~s. However, since there might be a (see Sec. \ref{sec:state2discussion}).

%than the result obtained from the fit to the $\alpha$-$\gamma$ events

\section{\label{sec:discussion}Discussion}

\subsection{\label{sec:state2discussion}$\alpha$-decay of the ($2^-$) state}

Because of the limited $\alpha$-energy resolution and the inability to extract the $b_\alpha$ value from the current data set, it was not possible to determine the reduced $\alpha$-decay widths. Nevertheless, some conclusions can be drawn from the presented results. 

The population of the low-spin states in $^{182}$Au is consistent with the pattern observed in the $\alpha$-decay studies of the neighboring isotope $^{184}$Tl, where the low-spin state has a higher $\alpha$-decay branching than the high-spin state. In $^{184}$Tl, $b_\alpha(2^-) = 1.22(30)\%$, compared to 0.47(6)\% for the $7^{(+)}$ state and 0.089(19)\% for the $10^{(-)}$ state \cite{CVanBeveren2016}. 

In-beam studies \cite{Zhang2002} revealed the existence of two bands in $^{182}$Au built on top of isomeric states with proposed spins and parities of ($6^+$) and ($10^-$). Thus, it cannot be excluded that the 10$^{(-)}$ state in $^{186}$Tl is $\alpha$ decaying directly to the ($10^-$) isomer in $^{182}$Au, without emission of a $\gamma$ ray. 

In the case of the possible $\alpha$ decay of the $^{186}$Tl($7^{(+)}$), it could decay to the $^{182}$Au($7^+$) state and then deexcite by emission of a 104 keV $\gamma$ ray to the ($6^+$) isomeric state. However, a $\gamma$ ray with the same energy is emitted from the decay of the 129 keV state in $^{182}$Au, known from the $\beta$-decay study of $^{182}$Hg \cite{Ibrahim2001}. The $\gamma$-intensity ratio of the 104 keV to 129 keV obtained from this work ($0.24^{+0.05}_{-0.04}$) is in agreement with the ratio extracted from the $\beta$-decay study ($0.23(2)$ \cite{Ibrahim2001}) and suggests that the majority of the 104 keV transition originates from the $\alpha$-decay of the $^{186}$Tl($2^-$) state. However, it should be noted that the 104 keV transition (from the 129 keV level to the 25 keV level) is known to be $E1$ \cite{Ibrahim2001} with a total theoretical conversion coefficient $\alpha_{tot}(E1)=0.376(11)$ \cite{Kibedi2008}, while the 104 keV $(7^+) \rightarrow (6^+)$ transition would be a mixed $M1$/$E2$ transition with $\alpha_{tot}(E2) = 4.42(19)$ and  $\alpha_{tot}(M1) = 6.49(21)$ \cite{Kibedi2008}. Thus, it cannot be excluded that the $^{186}$Tl($7^{(+)}$) state has an $\alpha$-decay branch. This would be consistent with the observed difference between half-lives obtained from the $\alpha$ particles and $\alpha$-$\gamma$ events, however, the 5(7)\% excess intensity of $\alpha$-$\gamma$ events compared to the single-$\alpha$ events indicates that this branching should be small.

%... that the 104 keV transition ( from 129 keV  to 25 keV level) is known to be ...

\subsection{\label{sec:state10discussion}Decay of the $10^{(-)}$ state}

\begin{figure*}
\includegraphics[width=\textwidth]{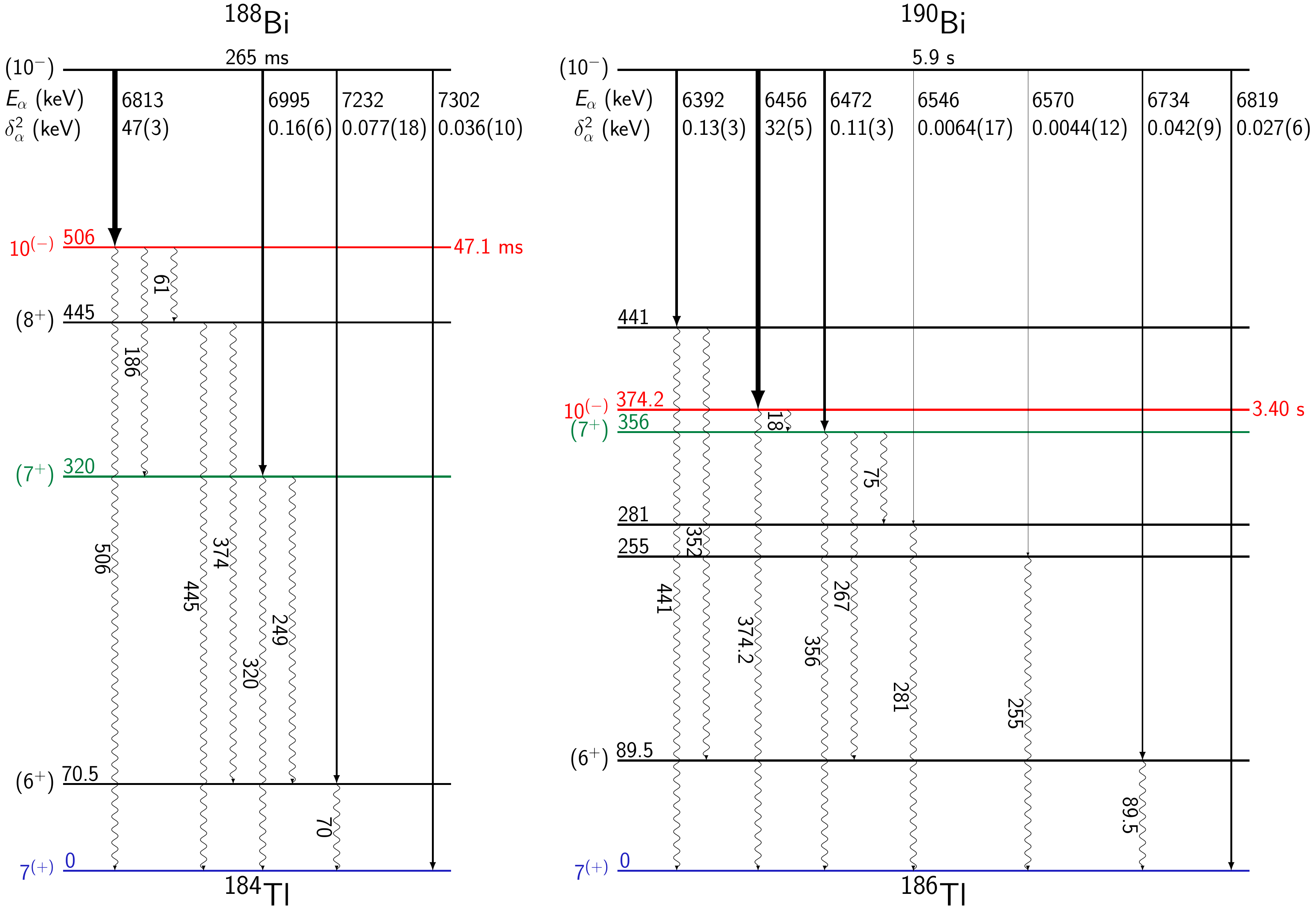}
\caption{\label{fig:systematics}(Color online) Systematic information from the $\alpha$- and isomeric decay studies in $^{184,186}$Tl. Next to each $\alpha$ transition, the energies and the reduced decay widths calculated using the Rasmussen formalism \cite{Rasmussen1959} are presented. The color code represents the main underlying proton single particle configuration as interpreted from magnetic moment measurements and systematics configurations \cite{Barzakh2013,VanBeveren2015,Barzakh2017} with the states involving the $\pi s_{1/2}$ orbital plotted in blue, the $\pi d_{3/2}$ orbital plotted in green, and the $\pi h_{9/2}$ orbital plotted in red. The energies of the excited states in thallium isotopes are shifted to set the $7^+_1$ states at 0. Experimental data are taken from our analysis and Refs. \cite{Andreyev2003,Barzakh2013,VanBeveren2015,Barzakh2017}}
\end{figure*}

\begin{table}
\caption{\label{tab:retardation}
The retardation factors for $M2$ and $E3$ transitions in $^{184,186,188}$Tl isotopes. Data for $^{186}$Tl are extracted from this work, while for the other isotopes are taken from Refs. \cite{VanBeveren2015,Kondev2018}. The retardation factors for $^{186}$Tl should be treated as a lower limit since the $\beta$-decay branching ratio is given only as a lower limit.}
\begin{ruledtabular}
\begin{tabular}{cccc}
Isotope 		& $E_{\gamma}$ (keV)	& Multipolarity	& $F_w$ 		\\\hline	
$^{184}$Tl 	& 61 					& $M2$			& $3.84(16) \times 10^4$ 	\\	
$^{184}$Tl 	& 186	 				& $E3$				& 486(115) 	\\
$^{184}$Tl 	& 506					& $E3$				& 984(38) 	\\
$^{186}$Tl 	& 18.2(4) 				& $E3$				& 1053(347) 	\\
$^{186}$Tl 	& 374.2(1)			& $E3$				& 7934(215) 	\\
$^{188}$Tl 	& 268.8				& $M2$			& $1.93(19) \times 10^5$ 	\\
\end{tabular}
\end{ruledtabular}
\end{table}

The isomeric character of the 374 keV state suggests that $M2$ or $E3$ are possible multipolarities for the 18 keV transition. For any higher multipolarity, the intensity of the 18 keV transition would be negligible compared to the 374 keV $E3$ transition. Retardation factors $F_w$, which are defined as a ratio of the Weisskopf estimate and the experimental transition strength, favor $E3$ multipolarity for the 18 keV compared to $M2$. The latter would lead to $F_w = 3.5(10) \times 10^7$ which is at least two orders of magnitude higher compared to the similar transitions in the neighboring isotopes (see Table \ref{tab:retardation}). Thus, the 18 keV transition is proposed to have multipolarity $E3$ and, consequently, the spin ($7^+$) has been tentatively assigned to the state at $x+356$ keV. This spin assignment is also supported by the $\alpha$-decay study of $^{188,190}$Bi to $^{184,186}$Tl \cite{Andreyev2003}. A reduced $\alpha$-decay width to the $x+356$ keV level in $^{186}$Tl ($\delta^2_\alpha = 0.11(3)$ keV) obtained in Ref. \cite{Andreyev2003} is similar to the reduced $\alpha$-decay width feeding the 320 keV level in $^{184}$Tl ($\delta^2_\alpha = 0.16(6)$ keV), which has been proposed to be the ($7^+_2$) state, see Fig. \ref{fig:systematics}.

%The isomeric character of the 374 keV state suggests that $M2$ or $E3$ are possible multipolarities for the 18 keV transition. For any higher multipolarity, the intensity of the 18 keV transition would be negligible compared to the 374 keV $E3$ transition. The 18 keV $E3$ transition is more favored than the $M2$ transition, as the latter would lead to high retardation factor of $F_w = 3.5(10) \times 10^7$, which is at least two orders of magnitude higher compared to the similar transitions in the neighboring isotopes (see Table \ref{tab:retardation}). Thus, the 18 keV transition is proposed to have multipolarity $E3$ and, consequently, the spin ($7^+$) has been tentatively assigned to the state at $x+356$ keV. This spin assignment is also supported by the $\alpha$-decay study of $^{188,190}$Bi to $^{184,186}$Tl \cite{Andreyev2003}. A reduced $\alpha$-decay width to the $x+356$ keV level in $^{186}$Tl ($\delta^2_\alpha = 0.11(3)$ keV) obtained in Ref. \cite{Andreyev2003} is similar to the reduced $\alpha$-decay width feeding the 320 keV level in $^{184}$Tl ($\delta^2_\alpha = 0.16(6)$ keV), which has been proposed to be the ($7^+_2$) state, see Fig. \ref{fig:systematics}.

%$10^- \rightarrow 7^+$ $E3$

The measurements of the magnetic moments indicated that $^{186}$Tl($7^+_1$) is dominated by the $[\pi s_{1/2} \otimes \nu i_{13/2}]_{7^+}$ configuration, while $^{184}$Tl($7^+_1$) should have a $\approx$20\% admixture from a $[\pi d_{3/2} \otimes \nu i_{13/2}]_{7^+}$ configuration \cite{Schuessler1995,Barzakh2017}. This admixture can explain the difference in the retardation factors of the 506 keV and 374 keV $E3$ transitions (Table \ref{tab:retardation}). In both $^{184}$Tl and $^{186}$Tl the dominant configuration of the $10^-$ intruder state was determined from the magnetic moment measurements to be $[\pi h_{9/2} \otimes \nu i_{13/2}]_{10^-}$ \cite{Barzakh2013,Barzakh2017} and from a single-particle approach, the transition between the $\pi h_{9/2}$ and $\pi d_{3/2}$ requires a smaller angular momentum change ($\Delta\ell = 3$) compared to the transition between the $\pi h_{9/2}$ and $\pi s_{1/2}$ ($\Delta\ell = 5$). This change was proposed in Ref. \cite{VanBeveren2015} to be the reason for the 374 keV transition in $^{186}$Tl being retarded by an order of magnitude more, relative to the 506 keV transition in $^{184}$Tl. In contrast, the retardation of the $10^- \rightarrow 7^+_2$ transitions in $^{184}$Tl and $^{186}$Tl are within the factor of two similar (note the large experimental uncertainties). This can be explained as due to a substantial contribution of the proton $d_{3/2}$ component in these $7^+_2$ states. 

%The almost an order of magnitude more retarded transition to the $7^+_1$ state in $^{186}$Tl compared to $^{184}$Tl was explained in this way in Ref. \cite{VanBeveren2015}. 

%This change was proposed to be the reason for the transition to the 7 state being retarded by an order of magnitude, relative to...

\section{\label{sec:conclusions}Conclusions}

The decay of the long-lived states in $^{186}$Tl was studied at ISOLDE, CERN. The existence of the ($2^-$) state was confirmed and its half-life of $3.4^{+0.5}_{-0.4}$~s was determined for the first time through the observation of its $\alpha$ decay. Six $\gamma$-ray transitions were measured in coincidence with the $\alpha$ decay of $^{186}$Tl, which allowed the construction of an $\alpha$-decay scheme feeding the levels in $^{182}$Au. 

Based on the $\alpha$-decay energy, the $7^{(+)}$ state in $^{186}$Tl was positioned 77(56)~keV above the ($2^-$) ground state and the ($10^-$) level in $^{190}$Bi was placed 182(57)~keV above the ($3^+$) ground state.  The decay scheme of the $^{186}$Tl($10^{(-)}$) state was extended, the half-life of $3.40(9)$~s was measured and a limit for the $\beta$-decay branching ratio has been deduced. The more detailed internal decay pattern observed for the $10^{(-)}$ state is consistent with the `intruder' origin of the isomer and follows a clear trend throughout the thallium isotopic chain. Based on the $E3$ retardation factors for the $10^{(-)} \rightarrow 7^{(+)}$ transitions, as well as additional information from bismuth $\alpha$-decay studies, the $x+356$ keV state was tentatively assigned with ($7^+$) spin. A comparison with a neighboring isotope of $^{184}$Tl indicated significant differences in the structure of the $7^+$ states. 

\begin{acknowledgments}

We acknowledge the support of the ISOLDE Collaboration and technical teams. This project has received funding from the European Union’s Horizon 2020 research and innovation programme grant agreements No. 654002 (ENSAR2) and No. 665779. This work has been funded by FWO-Vlaanderen (Belgium), by GOA/2015/010 (BOF KU Leuven), by the Interuniversity Attraction Poles Programme initiated by the Belgian Science Policy Office (BriX network P7/12), by the Slovak Research and Development Agency (Contract No. APVV-18-0268), by the Slovak Grant Agency VEGA (Contract No. 1/0532/17), by the Spanish government under contracts FPA2015-64969-P, FPA2015-65035-P, FPA2017-87568-P, FPA2017-83946-C2-1-P and RTI2018-098868-B-I00 (FEDER, UE), by Technology Facilities Council (STFC) of the UK Grant No. ST/R004056/1, by the German BMBF under contract 05P18PKCIA, by the Romanian IFA project CERN-RO/ISOLDE, by the National Science Centre (Poland), Grant No. 2015/18/M/ST2/00523 and by the Academy of Finland (Finland) Grant No. 307685.

\end{acknowledgments}

%\appendix
%\section{\label{apx:511intensity}Intensity of the 511 keV transition in the $^{66}$Fe to $^{66}$Co decay}

\bibliographystyle{apsrev4-1}
\bibliography{186Tlref,other}

\begin{thebibliography}{36}%
\makeatletter
\providecommand \@ifxundefined [1]{%
 \@ifx{#1\undefined}
}%
\providecommand \@ifnum [1]{%
 \ifnum #1\expandafter \@firstoftwo
 \else \expandafter \@secondoftwo
 \fi
}%
\providecommand \@ifx [1]{%
 \ifx #1\expandafter \@firstoftwo
 \else \expandafter \@secondoftwo
 \fi
}%
\providecommand \natexlab [1]{#1}%
\providecommand \enquote  [1]{``#1''}%
\providecommand \bibnamefont  [1]{#1}%
\providecommand \bibfnamefont [1]{#1}%
\providecommand \citenamefont [1]{#1}%
\providecommand \href@noop [0]{\@secondoftwo}%
\providecommand \href [0]{\begingroup \@sanitize@url \@href}%
\providecommand \@href[1]{\@@startlink{#1}\@@href}%
\providecommand \@@href[1]{\endgroup#1\@@endlink}%
\providecommand \@sanitize@url [0]{\catcode `\\12\catcode `\$12\catcode
  `\&12\catcode `\#12\catcode `\^12\catcode `\_12\catcode `\%12\relax}%
\providecommand \@@startlink[1]{}%
\providecommand \@@endlink[0]{}%
\providecommand \url  [0]{\begingroup\@sanitize@url \@url }%
\providecommand \@url [1]{\endgroup\@href {#1}{\urlprefix }}%
\providecommand \urlprefix  [0]{URL }%
\providecommand \Eprint [0]{\href }%
\providecommand \doibase [0]{http://dx.doi.org/}%
\providecommand \selectlanguage [0]{\@gobble}%
\providecommand \bibinfo  [0]{\@secondoftwo}%
\providecommand \bibfield  [0]{\@secondoftwo}%
\providecommand \translation [1]{[#1]}%
\providecommand \BibitemOpen [0]{}%
\providecommand \bibitemStop [0]{}%
\providecommand \bibitemNoStop [0]{.\EOS\space}%
\providecommand \EOS [0]{\spacefactor3000\relax}%
\providecommand \BibitemShut  [1]{\csname bibitem#1\endcsname}%
\let\auto@bib@innerbib\@empty
%</preamble>
\bibitem [{\citenamefont {Heyde}\ and\ \citenamefont {Wood}(2011)}]{Heyde2011}%
  \BibitemOpen
  \bibfield  {author} {\bibinfo {author} {\bibfnamefont {K.}~\bibnamefont
  {Heyde}}\ and\ \bibinfo {author} {\bibfnamefont {J.~L.}\ \bibnamefont
  {Wood}},\ }\href {\doibase 10.1103/RevModPhys.83.1467} {\bibfield  {journal}
  {\bibinfo  {journal} {Reviews of Modern Physics}\ }\textbf {\bibinfo {volume}
  {83}},\ \bibinfo {pages} {1467} (\bibinfo {year} {2011})}\BibitemShut
  {NoStop}%
\bibitem [{\citenamefont {Barzakh}\ \emph {et~al.}(2013)\citenamefont
  {Barzakh}, \citenamefont {Batist}, \citenamefont {Fedorov}, \citenamefont
  {Ivanov}, \citenamefont {Mezilev}, \citenamefont {Molkanov}, \citenamefont
  {Moroz}, \citenamefont {Orlov}, \citenamefont {Panteleev},\ and\
  \citenamefont {Volkov}}]{Barzakh2013}%
  \BibitemOpen
  \bibfield  {author} {\bibinfo {author} {\bibfnamefont {A.~E.}\ \bibnamefont
  {Barzakh}}, \bibinfo {author} {\bibfnamefont {L.~K.}\ \bibnamefont {Batist}},
  \bibinfo {author} {\bibfnamefont {D.~V.}\ \bibnamefont {Fedorov}}, \bibinfo
  {author} {\bibfnamefont {V.~S.}\ \bibnamefont {Ivanov}}, \bibinfo {author}
  {\bibfnamefont {K.~A.}\ \bibnamefont {Mezilev}}, \bibinfo {author}
  {\bibfnamefont {P.~L.}\ \bibnamefont {Molkanov}}, \bibinfo {author}
  {\bibfnamefont {F.~V.}\ \bibnamefont {Moroz}}, \bibinfo {author}
  {\bibfnamefont {S.~Y.}\ \bibnamefont {Orlov}}, \bibinfo {author}
  {\bibfnamefont {V.~N.}\ \bibnamefont {Panteleev}}, \ and\ \bibinfo {author}
  {\bibfnamefont {Y.~M.}\ \bibnamefont {Volkov}},\ }\href {\doibase
  10.1103/PhysRevC.88.024315} {\bibfield  {journal} {\bibinfo  {journal}
  {Physical Review C}\ }\textbf {\bibinfo {volume} {88}},\ \bibinfo {pages}
  {024315} (\bibinfo {year} {2013})}\BibitemShut {NoStop}%
\bibitem [{\citenamefont {Barzakh}\ \emph {et~al.}(2017)\citenamefont
  {Barzakh}, \citenamefont {Andreyev}, \citenamefont {Cocolios}, \citenamefont
  {de~Groote}, \citenamefont {Fedorov}, \citenamefont {Fedosseev},
  \citenamefont {Ferrer}, \citenamefont {Fink}, \citenamefont {Ghys},
  \citenamefont {Huyse}, \citenamefont {K{\"{o}}ster}, \citenamefont {Lane},
  \citenamefont {Liberati}, \citenamefont {Lynch}, \citenamefont {Marsh},
  \citenamefont {Molkanov}, \citenamefont {Procter}, \citenamefont {Rapisarda},
  \citenamefont {Rothe}, \citenamefont {Sandhu}, \citenamefont {Seliverstov},
  \citenamefont {Sj{\"{o}}din}, \citenamefont {{Van Beveren}}, \citenamefont
  {{Van Duppen}}, \citenamefont {Venhart},\ and\ \citenamefont
  {Veselsk{\'{y}}}}]{Barzakh2017}%
  \BibitemOpen
  \bibfield  {author} {\bibinfo {author} {\bibfnamefont {A.~E.}\ \bibnamefont
  {Barzakh}}, \bibinfo {author} {\bibfnamefont {A.~N.}\ \bibnamefont
  {Andreyev}}, \bibinfo {author} {\bibfnamefont {T.~E.}\ \bibnamefont
  {Cocolios}}, \bibinfo {author} {\bibfnamefont {R.~P.}\ \bibnamefont
  {de~Groote}}, \bibinfo {author} {\bibfnamefont {D.~V.}\ \bibnamefont
  {Fedorov}}, \bibinfo {author} {\bibfnamefont {V.~N.}\ \bibnamefont
  {Fedosseev}}, \bibinfo {author} {\bibfnamefont {R.}~\bibnamefont {Ferrer}},
  \bibinfo {author} {\bibfnamefont {D.~A.}\ \bibnamefont {Fink}}, \bibinfo
  {author} {\bibfnamefont {L.}~\bibnamefont {Ghys}}, \bibinfo {author}
  {\bibfnamefont {M.}~\bibnamefont {Huyse}}, \bibinfo {author} {\bibfnamefont
  {U.}~\bibnamefont {K{\"{o}}ster}}, \bibinfo {author} {\bibfnamefont
  {J.}~\bibnamefont {Lane}}, \bibinfo {author} {\bibfnamefont {V.}~\bibnamefont
  {Liberati}}, \bibinfo {author} {\bibfnamefont {K.~M.}\ \bibnamefont {Lynch}},
  \bibinfo {author} {\bibfnamefont {B.~A.}\ \bibnamefont {Marsh}}, \bibinfo
  {author} {\bibfnamefont {P.~L.}\ \bibnamefont {Molkanov}}, \bibinfo {author}
  {\bibfnamefont {T.~J.}\ \bibnamefont {Procter}}, \bibinfo {author}
  {\bibfnamefont {E.}~\bibnamefont {Rapisarda}}, \bibinfo {author}
  {\bibfnamefont {S.}~\bibnamefont {Rothe}}, \bibinfo {author} {\bibfnamefont
  {K.}~\bibnamefont {Sandhu}}, \bibinfo {author} {\bibfnamefont {M.~D.}\
  \bibnamefont {Seliverstov}}, \bibinfo {author} {\bibfnamefont {A.~M.}\
  \bibnamefont {Sj{\"{o}}din}}, \bibinfo {author} {\bibfnamefont
  {C.}~\bibnamefont {{Van Beveren}}}, \bibinfo {author} {\bibfnamefont
  {P.}~\bibnamefont {{Van Duppen}}}, \bibinfo {author} {\bibfnamefont
  {M.}~\bibnamefont {Venhart}}, \ and\ \bibinfo {author} {\bibfnamefont
  {M.}~\bibnamefont {Veselsk{\'{y}}}},\ }\href {\doibase
  10.1103/PhysRevC.95.014324} {\bibfield  {journal} {\bibinfo  {journal}
  {Physical Review C}\ }\textbf {\bibinfo {volume} {95}},\ \bibinfo {pages}
  {014324} (\bibinfo {year} {2017})}\BibitemShut {NoStop}%
\bibitem [{\citenamefont {Marsh}\ \emph {et~al.}(2018)\citenamefont {Marsh},
  \citenamefont {{Day Goodacre}}, \citenamefont {Sels}, \citenamefont
  {Tsunoda}, \citenamefont {Andel}, \citenamefont {Andreyev}, \citenamefont
  {Althubiti}, \citenamefont {Atanasov}, \citenamefont {Barzakh}, \citenamefont
  {Billowes}, \citenamefont {Blaum}, \citenamefont {Cocolios}, \citenamefont
  {Cubiss}, \citenamefont {Dobaczewski}, \citenamefont {Farooq-Smith},
  \citenamefont {Fedorov}, \citenamefont {Fedosseev}, \citenamefont {Flanagan},
  \citenamefont {Gaffney}, \citenamefont {Ghys}, \citenamefont {Huyse},
  \citenamefont {Kreim}, \citenamefont {Lunney}, \citenamefont {Lynch},
  \citenamefont {Manea}, \citenamefont {{Martinez Palenzuela}}, \citenamefont
  {Molkanov}, \citenamefont {Otsuka}, \citenamefont {Pastore}, \citenamefont
  {Rosenbusch}, \citenamefont {Rossel}, \citenamefont {Rothe}, \citenamefont
  {Schweikhard}, \citenamefont {Seliverstov}, \citenamefont {Spagnoletti},
  \citenamefont {{Van Beveren}}, \citenamefont {{Van Duppen}}, \citenamefont
  {Veinhard}, \citenamefont {Verstraelen}, \citenamefont {Welker},
  \citenamefont {Wendt}, \citenamefont {Wienholtz}, \citenamefont {Wolf},
  \citenamefont {Zadvornaya},\ and\ \citenamefont {Zuber}}]{Marsh2018}%
  \BibitemOpen
  \bibfield  {author} {\bibinfo {author} {\bibfnamefont {B.~A.}\ \bibnamefont
  {Marsh}}, \bibinfo {author} {\bibfnamefont {T.}~\bibnamefont {{Day
  Goodacre}}}, \bibinfo {author} {\bibfnamefont {S.}~\bibnamefont {Sels}},
  \bibinfo {author} {\bibfnamefont {Y.}~\bibnamefont {Tsunoda}}, \bibinfo
  {author} {\bibfnamefont {B.}~\bibnamefont {Andel}}, \bibinfo {author}
  {\bibfnamefont {A.~N.}\ \bibnamefont {Andreyev}}, \bibinfo {author}
  {\bibfnamefont {N.~A.}\ \bibnamefont {Althubiti}}, \bibinfo {author}
  {\bibfnamefont {D.}~\bibnamefont {Atanasov}}, \bibinfo {author}
  {\bibfnamefont {A.~E.}\ \bibnamefont {Barzakh}}, \bibinfo {author}
  {\bibfnamefont {J.}~\bibnamefont {Billowes}}, \bibinfo {author}
  {\bibfnamefont {K.}~\bibnamefont {Blaum}}, \bibinfo {author} {\bibfnamefont
  {T.~E.}\ \bibnamefont {Cocolios}}, \bibinfo {author} {\bibfnamefont {J.~G.}\
  \bibnamefont {Cubiss}}, \bibinfo {author} {\bibfnamefont {J.}~\bibnamefont
  {Dobaczewski}}, \bibinfo {author} {\bibfnamefont {G.~J.}\ \bibnamefont
  {Farooq-Smith}}, \bibinfo {author} {\bibfnamefont {D.~V.}\ \bibnamefont
  {Fedorov}}, \bibinfo {author} {\bibfnamefont {V.~N.}\ \bibnamefont
  {Fedosseev}}, \bibinfo {author} {\bibfnamefont {K.~T.}\ \bibnamefont
  {Flanagan}}, \bibinfo {author} {\bibfnamefont {L.~P.}\ \bibnamefont
  {Gaffney}}, \bibinfo {author} {\bibfnamefont {L.}~\bibnamefont {Ghys}},
  \bibinfo {author} {\bibfnamefont {M.}~\bibnamefont {Huyse}}, \bibinfo
  {author} {\bibfnamefont {S.}~\bibnamefont {Kreim}}, \bibinfo {author}
  {\bibfnamefont {D.}~\bibnamefont {Lunney}}, \bibinfo {author} {\bibfnamefont
  {K.~M.}\ \bibnamefont {Lynch}}, \bibinfo {author} {\bibfnamefont
  {V.}~\bibnamefont {Manea}}, \bibinfo {author} {\bibfnamefont
  {Y.}~\bibnamefont {{Martinez Palenzuela}}}, \bibinfo {author} {\bibfnamefont
  {P.~L.}\ \bibnamefont {Molkanov}}, \bibinfo {author} {\bibfnamefont
  {T.}~\bibnamefont {Otsuka}}, \bibinfo {author} {\bibfnamefont
  {A.}~\bibnamefont {Pastore}}, \bibinfo {author} {\bibfnamefont
  {M.}~\bibnamefont {Rosenbusch}}, \bibinfo {author} {\bibfnamefont {R.~E.}\
  \bibnamefont {Rossel}}, \bibinfo {author} {\bibfnamefont {S.}~\bibnamefont
  {Rothe}}, \bibinfo {author} {\bibfnamefont {L.}~\bibnamefont {Schweikhard}},
  \bibinfo {author} {\bibfnamefont {M.~D.}\ \bibnamefont {Seliverstov}},
  \bibinfo {author} {\bibfnamefont {P.}~\bibnamefont {Spagnoletti}}, \bibinfo
  {author} {\bibfnamefont {C.}~\bibnamefont {{Van Beveren}}}, \bibinfo {author}
  {\bibfnamefont {P.}~\bibnamefont {{Van Duppen}}}, \bibinfo {author}
  {\bibfnamefont {M.}~\bibnamefont {Veinhard}}, \bibinfo {author}
  {\bibfnamefont {E.}~\bibnamefont {Verstraelen}}, \bibinfo {author}
  {\bibfnamefont {A.}~\bibnamefont {Welker}}, \bibinfo {author} {\bibfnamefont
  {K.}~\bibnamefont {Wendt}}, \bibinfo {author} {\bibfnamefont
  {F.}~\bibnamefont {Wienholtz}}, \bibinfo {author} {\bibfnamefont {R.~N.}\
  \bibnamefont {Wolf}}, \bibinfo {author} {\bibfnamefont {A.}~\bibnamefont
  {Zadvornaya}}, \ and\ \bibinfo {author} {\bibfnamefont {K.}~\bibnamefont
  {Zuber}},\ }\href {\doibase 10.1038/s41567-018-0292-8} {\bibfield  {journal}
  {\bibinfo  {journal} {Nature Physics}\ }\textbf {\bibinfo {volume} {14}},\
  \bibinfo {pages} {1163} (\bibinfo {year} {2018})}\BibitemShut {NoStop}%
\bibitem [{\citenamefont {Sels}\ \emph {et~al.}(2019)\citenamefont {Sels},
  \citenamefont {{Day Goodacre}}, \citenamefont {Marsh}, \citenamefont
  {Pastore}, \citenamefont {Ryssens}, \citenamefont {Tsunoda}, \citenamefont
  {Althubiti}, \citenamefont {Andel}, \citenamefont {Andreyev}, \citenamefont
  {Atanasov}, \citenamefont {Barzakh}, \citenamefont {Bender}, \citenamefont
  {Billowes}, \citenamefont {Blaum}, \citenamefont {Cocolios}, \citenamefont
  {Cubiss}, \citenamefont {Dobaczewski}, \citenamefont {Farooq-Smith},
  \citenamefont {Fedorov}, \citenamefont {Fedosseev}, \citenamefont {Flanagan},
  \citenamefont {Gaffney}, \citenamefont {Ghys}, \citenamefont {Heenen},
  \citenamefont {Huyse}, \citenamefont {Kreim}, \citenamefont {Lunney},
  \citenamefont {Lynch}, \citenamefont {Manea}, \citenamefont {{Martinez
  Palenzuela}}, \citenamefont {Medonca}, \citenamefont {Molkanov},
  \citenamefont {Otsuka}, \citenamefont {Ramos}, \citenamefont {Rossel},
  \citenamefont {Rothe}, \citenamefont {Schweikhard}, \citenamefont
  {Seliverstov}, \citenamefont {Spagnoletti}, \citenamefont {{Van Beveren}},
  \citenamefont {{Van Duppen}}, \citenamefont {Veinhard}, \citenamefont
  {Verstraelen}, \citenamefont {Welker}, \citenamefont {Wendt}, \citenamefont
  {Wienholtz}, \citenamefont {Wolf},\ and\ \citenamefont
  {Zadvornaya}}]{Sels2019}%
  \BibitemOpen
  \bibfield  {author} {\bibinfo {author} {\bibfnamefont {S.}~\bibnamefont
  {Sels}}, \bibinfo {author} {\bibfnamefont {T.}~\bibnamefont {{Day
  Goodacre}}}, \bibinfo {author} {\bibfnamefont {B.~A.}\ \bibnamefont {Marsh}},
  \bibinfo {author} {\bibfnamefont {A.}~\bibnamefont {Pastore}}, \bibinfo
  {author} {\bibfnamefont {W.}~\bibnamefont {Ryssens}}, \bibinfo {author}
  {\bibfnamefont {Y.}~\bibnamefont {Tsunoda}}, \bibinfo {author} {\bibfnamefont
  {N.}~\bibnamefont {Althubiti}}, \bibinfo {author} {\bibfnamefont
  {B.}~\bibnamefont {Andel}}, \bibinfo {author} {\bibfnamefont {A.~N.}\
  \bibnamefont {Andreyev}}, \bibinfo {author} {\bibfnamefont {D.}~\bibnamefont
  {Atanasov}}, \bibinfo {author} {\bibfnamefont {A.~E.}\ \bibnamefont
  {Barzakh}}, \bibinfo {author} {\bibfnamefont {M.}~\bibnamefont {Bender}},
  \bibinfo {author} {\bibfnamefont {J.}~\bibnamefont {Billowes}}, \bibinfo
  {author} {\bibfnamefont {K.}~\bibnamefont {Blaum}}, \bibinfo {author}
  {\bibfnamefont {T.~E.}\ \bibnamefont {Cocolios}}, \bibinfo {author}
  {\bibfnamefont {J.~G.}\ \bibnamefont {Cubiss}}, \bibinfo {author}
  {\bibfnamefont {J.}~\bibnamefont {Dobaczewski}}, \bibinfo {author}
  {\bibfnamefont {G.~J.}\ \bibnamefont {Farooq-Smith}}, \bibinfo {author}
  {\bibfnamefont {D.~V.}\ \bibnamefont {Fedorov}}, \bibinfo {author}
  {\bibfnamefont {V.~N.}\ \bibnamefont {Fedosseev}}, \bibinfo {author}
  {\bibfnamefont {K.~T.}\ \bibnamefont {Flanagan}}, \bibinfo {author}
  {\bibfnamefont {L.~P.}\ \bibnamefont {Gaffney}}, \bibinfo {author}
  {\bibfnamefont {L.}~\bibnamefont {Ghys}}, \bibinfo {author} {\bibfnamefont
  {P.-H.}\ \bibnamefont {Heenen}}, \bibinfo {author} {\bibfnamefont
  {M.}~\bibnamefont {Huyse}}, \bibinfo {author} {\bibfnamefont
  {S.}~\bibnamefont {Kreim}}, \bibinfo {author} {\bibfnamefont
  {D.}~\bibnamefont {Lunney}}, \bibinfo {author} {\bibfnamefont {K.~M.}\
  \bibnamefont {Lynch}}, \bibinfo {author} {\bibfnamefont {V.}~\bibnamefont
  {Manea}}, \bibinfo {author} {\bibfnamefont {Y.}~\bibnamefont {{Martinez
  Palenzuela}}}, \bibinfo {author} {\bibfnamefont {T.~M.}\ \bibnamefont
  {Medonca}}, \bibinfo {author} {\bibfnamefont {P.~L.}\ \bibnamefont
  {Molkanov}}, \bibinfo {author} {\bibfnamefont {T.}~\bibnamefont {Otsuka}},
  \bibinfo {author} {\bibfnamefont {J.~P.}\ \bibnamefont {Ramos}}, \bibinfo
  {author} {\bibfnamefont {R.~E.}\ \bibnamefont {Rossel}}, \bibinfo {author}
  {\bibfnamefont {S.}~\bibnamefont {Rothe}}, \bibinfo {author} {\bibfnamefont
  {L.}~\bibnamefont {Schweikhard}}, \bibinfo {author} {\bibfnamefont {M.~D.}\
  \bibnamefont {Seliverstov}}, \bibinfo {author} {\bibfnamefont
  {P.}~\bibnamefont {Spagnoletti}}, \bibinfo {author} {\bibfnamefont
  {C.}~\bibnamefont {{Van Beveren}}}, \bibinfo {author} {\bibfnamefont
  {P.}~\bibnamefont {{Van Duppen}}}, \bibinfo {author} {\bibfnamefont
  {M.}~\bibnamefont {Veinhard}}, \bibinfo {author} {\bibfnamefont
  {E.}~\bibnamefont {Verstraelen}}, \bibinfo {author} {\bibfnamefont
  {A.}~\bibnamefont {Welker}}, \bibinfo {author} {\bibfnamefont
  {K.}~\bibnamefont {Wendt}}, \bibinfo {author} {\bibfnamefont
  {F.}~\bibnamefont {Wienholtz}}, \bibinfo {author} {\bibfnamefont {R.~N.}\
  \bibnamefont {Wolf}}, \ and\ \bibinfo {author} {\bibfnamefont
  {A.}~\bibnamefont {Zadvornaya}},\ }\href {\doibase
  10.1103/PhysRevC.99.044306} {\bibfield  {journal} {\bibinfo  {journal}
  {Physical Review C}\ }\textbf {\bibinfo {volume} {99}},\ \bibinfo {pages}
  {044306} (\bibinfo {year} {2019})},\ \Eprint
  {http://arxiv.org/abs/1902.11211} {arXiv:1902.11211} \BibitemShut {NoStop}%
\bibitem [{\citenamefont {Andreyev}\ \emph
  {et~al.}(2003{\natexlab{a}})\citenamefont {Andreyev}, \citenamefont
  {Ackermann}, \citenamefont {Antalic}, \citenamefont {Boardman}, \citenamefont
  {Cagarda}, \citenamefont {Gerl}, \citenamefont {He{\ss}berger}, \citenamefont
  {Hofmann}, \citenamefont {Huyse}, \citenamefont {Karlgren}, \citenamefont
  {Keenan}, \citenamefont {Kettunen}, \citenamefont {Kleinb{\"{o}}hl},
  \citenamefont {Kindler}, \citenamefont {Kojouharov}, \citenamefont
  {Lavrentiev}, \citenamefont {O'Leary}, \citenamefont {Leino}, \citenamefont
  {Lommel}, \citenamefont {Matos}, \citenamefont {Moore}, \citenamefont
  {M{\"{u}}nzenberg}, \citenamefont {Page}, \citenamefont {Reshitko},
  \citenamefont {Saro}, \citenamefont {Schaffner}, \citenamefont {Schlegel},
  \citenamefont {Taylor}, \citenamefont {{Van de Vel}}, \citenamefont {{Van
  Duppen}}, \citenamefont {Weissman},\ and\ \citenamefont
  {Heyde}}]{Andreyev2003}%
  \BibitemOpen
  \bibfield  {author} {\bibinfo {author} {\bibfnamefont {A.~N.}\ \bibnamefont
  {Andreyev}}, \bibinfo {author} {\bibfnamefont {D.}~\bibnamefont {Ackermann}},
  \bibinfo {author} {\bibfnamefont {S.}~\bibnamefont {Antalic}}, \bibinfo
  {author} {\bibfnamefont {H.~J.}\ \bibnamefont {Boardman}}, \bibinfo {author}
  {\bibfnamefont {P.}~\bibnamefont {Cagarda}}, \bibinfo {author} {\bibfnamefont
  {J.}~\bibnamefont {Gerl}}, \bibinfo {author} {\bibfnamefont {F.~P.}\
  \bibnamefont {He{\ss}berger}}, \bibinfo {author} {\bibfnamefont
  {S.}~\bibnamefont {Hofmann}}, \bibinfo {author} {\bibfnamefont
  {M.}~\bibnamefont {Huyse}}, \bibinfo {author} {\bibfnamefont
  {D.}~\bibnamefont {Karlgren}}, \bibinfo {author} {\bibfnamefont
  {A.}~\bibnamefont {Keenan}}, \bibinfo {author} {\bibfnamefont
  {H.}~\bibnamefont {Kettunen}}, \bibinfo {author} {\bibfnamefont
  {A.}~\bibnamefont {Kleinb{\"{o}}hl}}, \bibinfo {author} {\bibfnamefont
  {B.}~\bibnamefont {Kindler}}, \bibinfo {author} {\bibfnamefont
  {I.}~\bibnamefont {Kojouharov}}, \bibinfo {author} {\bibfnamefont
  {A.}~\bibnamefont {Lavrentiev}}, \bibinfo {author} {\bibfnamefont {C.~D.}\
  \bibnamefont {O'Leary}}, \bibinfo {author} {\bibfnamefont {M.}~\bibnamefont
  {Leino}}, \bibinfo {author} {\bibfnamefont {B.}~\bibnamefont {Lommel}},
  \bibinfo {author} {\bibfnamefont {M.}~\bibnamefont {Matos}}, \bibinfo
  {author} {\bibfnamefont {C.~J.}\ \bibnamefont {Moore}}, \bibinfo {author}
  {\bibfnamefont {G.}~\bibnamefont {M{\"{u}}nzenberg}}, \bibinfo {author}
  {\bibfnamefont {R.~D.}\ \bibnamefont {Page}}, \bibinfo {author}
  {\bibfnamefont {S.}~\bibnamefont {Reshitko}}, \bibinfo {author}
  {\bibfnamefont {S.}~\bibnamefont {Saro}}, \bibinfo {author} {\bibfnamefont
  {H.}~\bibnamefont {Schaffner}}, \bibinfo {author} {\bibfnamefont
  {C.}~\bibnamefont {Schlegel}}, \bibinfo {author} {\bibfnamefont {M.~J.}\
  \bibnamefont {Taylor}}, \bibinfo {author} {\bibfnamefont {K.}~\bibnamefont
  {{Van de Vel}}}, \bibinfo {author} {\bibfnamefont {P.}~\bibnamefont {{Van
  Duppen}}}, \bibinfo {author} {\bibfnamefont {L.}~\bibnamefont {Weissman}}, \
  and\ \bibinfo {author} {\bibfnamefont {K.}~\bibnamefont {Heyde}},\ }\href
  {\doibase 10.1140/epja/i2003-10050-2} {\bibfield  {journal} {\bibinfo
  {journal} {The European Physical Journal A}\ }\textbf {\bibinfo {volume}
  {18}},\ \bibinfo {pages} {39} (\bibinfo {year}
  {2003}{\natexlab{a}})}\BibitemShut {NoStop}%
\bibitem [{\citenamefont {Andreyev}\ \emph
  {et~al.}(2003{\natexlab{b}})\citenamefont {Andreyev}, \citenamefont
  {Ackermann}, \citenamefont {He{\ss}berger}, \citenamefont {Hofmann},
  \citenamefont {Huyse}, \citenamefont {Kojouharov}, \citenamefont {Kindler},
  \citenamefont {Lommel}, \citenamefont {M{\"{u}}nzenberg}, \citenamefont
  {Page}, \citenamefont {{Van de Vel}}, \citenamefont {{Van Duppen}},\ and\
  \citenamefont {Heyde}}]{Andreyev2003a}%
  \BibitemOpen
  \bibfield  {author} {\bibinfo {author} {\bibfnamefont {A.~N.}\ \bibnamefont
  {Andreyev}}, \bibinfo {author} {\bibfnamefont {D.}~\bibnamefont {Ackermann}},
  \bibinfo {author} {\bibfnamefont {F.~P.}\ \bibnamefont {He{\ss}berger}},
  \bibinfo {author} {\bibfnamefont {S.}~\bibnamefont {Hofmann}}, \bibinfo
  {author} {\bibfnamefont {M.}~\bibnamefont {Huyse}}, \bibinfo {author}
  {\bibfnamefont {I.}~\bibnamefont {Kojouharov}}, \bibinfo {author}
  {\bibfnamefont {B.}~\bibnamefont {Kindler}}, \bibinfo {author} {\bibfnamefont
  {B.}~\bibnamefont {Lommel}}, \bibinfo {author} {\bibfnamefont
  {G.}~\bibnamefont {M{\"{u}}nzenberg}}, \bibinfo {author} {\bibfnamefont
  {R.~D.}\ \bibnamefont {Page}}, \bibinfo {author} {\bibfnamefont
  {K.}~\bibnamefont {{Van de Vel}}}, \bibinfo {author} {\bibfnamefont
  {P.}~\bibnamefont {{Van Duppen}}}, \ and\ \bibinfo {author} {\bibfnamefont
  {K.}~\bibnamefont {Heyde}},\ }\href {\doibase 10.1140/epja/i2003-10051-1}
  {\bibfield  {journal} {\bibinfo  {journal} {The European Physical Journal A}\
  }\textbf {\bibinfo {volume} {18}},\ \bibinfo {pages} {55} (\bibinfo {year}
  {2003}{\natexlab{b}})}\BibitemShut {NoStop}%
\bibitem [{\citenamefont {{Van Beveren}}\ \emph {et~al.}(2015)\citenamefont
  {{Van Beveren}}, \citenamefont {Andreyev}, \citenamefont {Barzakh},
  \citenamefont {Cocolios}, \citenamefont {Fedorov}, \citenamefont {Fedosseev},
  \citenamefont {Ferrer}, \citenamefont {Huyse}, \citenamefont {K{\"{o}}ster},
  \citenamefont {Lane}, \citenamefont {Liberati}, \citenamefont {Lynch},
  \citenamefont {Marsh}, \citenamefont {Procter}, \citenamefont {Radulov},
  \citenamefont {Rapisarda}, \citenamefont {Sandhu}, \citenamefont
  {Seliverstov}, \citenamefont {{Van Duppen}}, \citenamefont {Venhart},\ and\
  \citenamefont {Veselsky}}]{VanBeveren2015}%
  \BibitemOpen
  \bibfield  {author} {\bibinfo {author} {\bibfnamefont {C.}~\bibnamefont {{Van
  Beveren}}}, \bibinfo {author} {\bibfnamefont {A.~N.}\ \bibnamefont
  {Andreyev}}, \bibinfo {author} {\bibfnamefont {A.~E.}\ \bibnamefont
  {Barzakh}}, \bibinfo {author} {\bibfnamefont {T.~E.}\ \bibnamefont
  {Cocolios}}, \bibinfo {author} {\bibfnamefont {D.}~\bibnamefont {Fedorov}},
  \bibinfo {author} {\bibfnamefont {V.~N.}\ \bibnamefont {Fedosseev}}, \bibinfo
  {author} {\bibfnamefont {R.}~\bibnamefont {Ferrer}}, \bibinfo {author}
  {\bibfnamefont {M.}~\bibnamefont {Huyse}}, \bibinfo {author} {\bibfnamefont
  {U.}~\bibnamefont {K{\"{o}}ster}}, \bibinfo {author} {\bibfnamefont
  {J.}~\bibnamefont {Lane}}, \bibinfo {author} {\bibfnamefont {V.}~\bibnamefont
  {Liberati}}, \bibinfo {author} {\bibfnamefont {K.~M.}\ \bibnamefont {Lynch}},
  \bibinfo {author} {\bibfnamefont {B.~A.}\ \bibnamefont {Marsh}}, \bibinfo
  {author} {\bibfnamefont {T.~J.}\ \bibnamefont {Procter}}, \bibinfo {author}
  {\bibfnamefont {D.}~\bibnamefont {Radulov}}, \bibinfo {author} {\bibfnamefont
  {E.}~\bibnamefont {Rapisarda}}, \bibinfo {author} {\bibfnamefont
  {K.}~\bibnamefont {Sandhu}}, \bibinfo {author} {\bibfnamefont {M.~D.}\
  \bibnamefont {Seliverstov}}, \bibinfo {author} {\bibfnamefont
  {P.}~\bibnamefont {{Van Duppen}}}, \bibinfo {author} {\bibfnamefont
  {M.}~\bibnamefont {Venhart}}, \ and\ \bibinfo {author} {\bibfnamefont
  {M.}~\bibnamefont {Veselsky}},\ }\href {\doibase 10.1103/PhysRevC.92.014325}
  {\bibfield  {journal} {\bibinfo  {journal} {Physical Review C}\ }\textbf
  {\bibinfo {volume} {92}},\ \bibinfo {pages} {014325} (\bibinfo {year}
  {2015})}\BibitemShut {NoStop}%
\bibitem [{\citenamefont {{Van Beveren}}\ \emph {et~al.}(2016)\citenamefont
  {{Van Beveren}}, \citenamefont {Andreyev}, \citenamefont {Barzakh},
  \citenamefont {Cocolios}, \citenamefont {{De Groote}}, \citenamefont
  {Fedorov}, \citenamefont {Fedosseev}, \citenamefont {Ferrer}, \citenamefont
  {Ghys}, \citenamefont {Huyse}, \citenamefont {K{\"{o}}ster}, \citenamefont
  {Lane}, \citenamefont {Liberati}, \citenamefont {Lynch}, \citenamefont
  {Marsh}, \citenamefont {Molkanov}, \citenamefont {Procter}, \citenamefont
  {Rapisarda}, \citenamefont {Sandhu}, \citenamefont {Seliverstov},
  \citenamefont {{Van Duppen}}, \citenamefont {Venhart},\ and\ \citenamefont
  {Veselsk{\'{y}}}}]{CVanBeveren2016}%
  \BibitemOpen
  \bibfield  {author} {\bibinfo {author} {\bibfnamefont {C.}~\bibnamefont {{Van
  Beveren}}}, \bibinfo {author} {\bibfnamefont {A.~N.}\ \bibnamefont
  {Andreyev}}, \bibinfo {author} {\bibfnamefont {A.~E.}\ \bibnamefont
  {Barzakh}}, \bibinfo {author} {\bibfnamefont {T.~E.}\ \bibnamefont
  {Cocolios}}, \bibinfo {author} {\bibfnamefont {R.~P.}\ \bibnamefont {{De
  Groote}}}, \bibinfo {author} {\bibfnamefont {D.}~\bibnamefont {Fedorov}},
  \bibinfo {author} {\bibfnamefont {V.~N.}\ \bibnamefont {Fedosseev}}, \bibinfo
  {author} {\bibfnamefont {R.}~\bibnamefont {Ferrer}}, \bibinfo {author}
  {\bibfnamefont {L.}~\bibnamefont {Ghys}}, \bibinfo {author} {\bibfnamefont
  {M.}~\bibnamefont {Huyse}}, \bibinfo {author} {\bibfnamefont
  {U.}~\bibnamefont {K{\"{o}}ster}}, \bibinfo {author} {\bibfnamefont
  {J.}~\bibnamefont {Lane}}, \bibinfo {author} {\bibfnamefont {V.}~\bibnamefont
  {Liberati}}, \bibinfo {author} {\bibfnamefont {K.~M.}\ \bibnamefont {Lynch}},
  \bibinfo {author} {\bibfnamefont {B.~A.}\ \bibnamefont {Marsh}}, \bibinfo
  {author} {\bibfnamefont {P.~L.}\ \bibnamefont {Molkanov}}, \bibinfo {author}
  {\bibfnamefont {T.~J.}\ \bibnamefont {Procter}}, \bibinfo {author}
  {\bibfnamefont {E.}~\bibnamefont {Rapisarda}}, \bibinfo {author}
  {\bibfnamefont {K.}~\bibnamefont {Sandhu}}, \bibinfo {author} {\bibfnamefont
  {M.~D.}\ \bibnamefont {Seliverstov}}, \bibinfo {author} {\bibfnamefont
  {P.}~\bibnamefont {{Van Duppen}}}, \bibinfo {author} {\bibfnamefont
  {M.}~\bibnamefont {Venhart}}, \ and\ \bibinfo {author} {\bibfnamefont
  {M.}~\bibnamefont {Veselsk{\'{y}}}},\ }\href {\doibase
  10.1088/0954-3899/43/2/025102} {\bibfield  {journal} {\bibinfo  {journal}
  {Journal of Physics G: Nuclear and Particle Physics}\ }\textbf {\bibinfo
  {volume} {43}},\ \bibinfo {pages} {025102} (\bibinfo {year}
  {2016})}\BibitemShut {NoStop}%
\bibitem [{\citenamefont {Rapisarda}\ \emph {et~al.}(2017)\citenamefont
  {Rapisarda}, \citenamefont {Andreyev}, \citenamefont {Antalic}, \citenamefont
  {Barzakh}, \citenamefont {Cocolios}, \citenamefont {Darby}, \citenamefont
  {{De Groote}}, \citenamefont {{De Witte}}, \citenamefont {Diriken},
  \citenamefont {Elseviers}, \citenamefont {Fedorov}, \citenamefont
  {Fedosseev}, \citenamefont {Ferrer}, \citenamefont {Huyse}, \citenamefont
  {Kalaninov{\'{a}}}, \citenamefont {K{\"{o}}ster}, \citenamefont {Lane},
  \citenamefont {Liberati}, \citenamefont {Lynch}, \citenamefont {Marsh},
  \citenamefont {Molkanov}, \citenamefont {Pauwels}, \citenamefont {Procter},
  \citenamefont {Radulov}, \citenamefont {Sandhu}, \citenamefont {Seliverstov},
  \citenamefont {{Van Beveren}}, \citenamefont {{Van den Bergh}}, \citenamefont
  {{Van Duppen}}, \citenamefont {Venhart}, \citenamefont {Veselsk{\'{y}}},\
  and\ \citenamefont {Wrzosek-Lipska}}]{Rapisarda2017}%
  \BibitemOpen
  \bibfield  {author} {\bibinfo {author} {\bibfnamefont {E.}~\bibnamefont
  {Rapisarda}}, \bibinfo {author} {\bibfnamefont {A.~N.}\ \bibnamefont
  {Andreyev}}, \bibinfo {author} {\bibfnamefont {S.}~\bibnamefont {Antalic}},
  \bibinfo {author} {\bibfnamefont {A.}~\bibnamefont {Barzakh}}, \bibinfo
  {author} {\bibfnamefont {T.~E.}\ \bibnamefont {Cocolios}}, \bibinfo {author}
  {\bibfnamefont {I.~G.}\ \bibnamefont {Darby}}, \bibinfo {author}
  {\bibfnamefont {R.}~\bibnamefont {{De Groote}}}, \bibinfo {author}
  {\bibfnamefont {H.}~\bibnamefont {{De Witte}}}, \bibinfo {author}
  {\bibfnamefont {J.}~\bibnamefont {Diriken}}, \bibinfo {author} {\bibfnamefont
  {J.}~\bibnamefont {Elseviers}}, \bibinfo {author} {\bibfnamefont
  {D.}~\bibnamefont {Fedorov}}, \bibinfo {author} {\bibfnamefont {V.~N.}\
  \bibnamefont {Fedosseev}}, \bibinfo {author} {\bibfnamefont {R.}~\bibnamefont
  {Ferrer}}, \bibinfo {author} {\bibfnamefont {M.}~\bibnamefont {Huyse}},
  \bibinfo {author} {\bibfnamefont {Z.}~\bibnamefont {Kalaninov{\'{a}}}},
  \bibinfo {author} {\bibfnamefont {U.}~\bibnamefont {K{\"{o}}ster}}, \bibinfo
  {author} {\bibfnamefont {J.}~\bibnamefont {Lane}}, \bibinfo {author}
  {\bibfnamefont {V.}~\bibnamefont {Liberati}}, \bibinfo {author}
  {\bibfnamefont {K.~M.}\ \bibnamefont {Lynch}}, \bibinfo {author}
  {\bibfnamefont {B.~A.}\ \bibnamefont {Marsh}}, \bibinfo {author}
  {\bibfnamefont {P.~L.}\ \bibnamefont {Molkanov}}, \bibinfo {author}
  {\bibfnamefont {D.}~\bibnamefont {Pauwels}}, \bibinfo {author} {\bibfnamefont
  {T.~J.}\ \bibnamefont {Procter}}, \bibinfo {author} {\bibfnamefont
  {D.}~\bibnamefont {Radulov}}, \bibinfo {author} {\bibfnamefont
  {K.}~\bibnamefont {Sandhu}}, \bibinfo {author} {\bibfnamefont {M.~D.}\
  \bibnamefont {Seliverstov}}, \bibinfo {author} {\bibfnamefont
  {C.}~\bibnamefont {{Van Beveren}}}, \bibinfo {author} {\bibfnamefont
  {P.}~\bibnamefont {{Van den Bergh}}}, \bibinfo {author} {\bibfnamefont
  {P.}~\bibnamefont {{Van Duppen}}}, \bibinfo {author} {\bibfnamefont
  {M.}~\bibnamefont {Venhart}}, \bibinfo {author} {\bibfnamefont
  {M.}~\bibnamefont {Veselsk{\'{y}}}}, \ and\ \bibinfo {author} {\bibfnamefont
  {K.}~\bibnamefont {Wrzosek-Lipska}},\ }\href {\doibase
  10.1088/1361-6471/aa6bb6} {\bibfield  {journal} {\bibinfo  {journal} {Journal
  of Physics G: Nuclear and Particle Physics}\ }\textbf {\bibinfo {volume}
  {44}},\ \bibinfo {pages} {074001} (\bibinfo {year} {2017})}\BibitemShut
  {NoStop}%
\bibitem [{\citenamefont {Wrzosek-Lipska}\ \emph {et~al.}(2019)\citenamefont
  {Wrzosek-Lipska}, \citenamefont {Rezynkina}, \citenamefont {Bree},
  \citenamefont {Zieli{\'{n}}ska}, \citenamefont {Gaffney}, \citenamefont
  {Petts}, \citenamefont {Andreyev}, \citenamefont {Bastin}, \citenamefont
  {Bender}, \citenamefont {Blazhev}, \citenamefont {Bruyneel}, \citenamefont
  {Butler}, \citenamefont {Carpenter}, \citenamefont {Cederk{\"{a}}ll},
  \citenamefont {Cl{\'{e}}ment}, \citenamefont {Cocolios}, \citenamefont
  {Deacon}, \citenamefont {Diriken}, \citenamefont {Ekstr{\"{o}}m},
  \citenamefont {Fitzpatrick}, \citenamefont {Fraile}, \citenamefont {Fransen},
  \citenamefont {Freeman}, \citenamefont {Garc{\'{i}}a-Ramos}, \citenamefont
  {Geibel}, \citenamefont {Gernh{\"{a}}user}, \citenamefont {Grahn},
  \citenamefont {Guttormsen}, \citenamefont {Hadinia}, \citenamefont
  {Hady{\'{n}}ska-Kl{\c{e}}k}, \citenamefont {Hass}, \citenamefont {Heenen},
  \citenamefont {Herzberg}, \citenamefont {Hess}, \citenamefont {Heyde},
  \citenamefont {Huyse}, \citenamefont {Ivanov}, \citenamefont {Jenkins},
  \citenamefont {Julin}, \citenamefont {Kesteloot}, \citenamefont
  {Kr{\"{o}}ll}, \citenamefont {Kr{\"{u}}cken}, \citenamefont {Larsen},
  \citenamefont {Lutter}, \citenamefont {Marley}, \citenamefont {Napiorkowski},
  \citenamefont {Orlandi}, \citenamefont {Page}, \citenamefont {Pakarinen},
  \citenamefont {Patronis}, \citenamefont {Peura}, \citenamefont {Piselli},
  \citenamefont {Pr{\'{o}}chniak}, \citenamefont {Rahkila}, \citenamefont
  {Rapisarda}, \citenamefont {Reiter}, \citenamefont {Robinson}, \citenamefont
  {Scheck}, \citenamefont {Siem}, \citenamefont {{Singh Chakkal}},
  \citenamefont {Smith}, \citenamefont {Srebrny}, \citenamefont {Stefanescu},
  \citenamefont {Tveten}, \citenamefont {{Van Duppen}}, \citenamefont {{Van de
  Walle}}, \citenamefont {Voulot}, \citenamefont {Warr}, \citenamefont
  {Wiens},\ and\ \citenamefont {Wood}}]{Wrzosek-Lipska2019}%
  \BibitemOpen
  \bibfield  {author} {\bibinfo {author} {\bibfnamefont {K.}~\bibnamefont
  {Wrzosek-Lipska}}, \bibinfo {author} {\bibfnamefont {K.}~\bibnamefont
  {Rezynkina}}, \bibinfo {author} {\bibfnamefont {N.}~\bibnamefont {Bree}},
  \bibinfo {author} {\bibfnamefont {M.}~\bibnamefont {Zieli{\'{n}}ska}},
  \bibinfo {author} {\bibfnamefont {L.~P.}\ \bibnamefont {Gaffney}}, \bibinfo
  {author} {\bibfnamefont {A.}~\bibnamefont {Petts}}, \bibinfo {author}
  {\bibfnamefont {A.}~\bibnamefont {Andreyev}}, \bibinfo {author}
  {\bibfnamefont {B.}~\bibnamefont {Bastin}}, \bibinfo {author} {\bibfnamefont
  {M.}~\bibnamefont {Bender}}, \bibinfo {author} {\bibfnamefont
  {A.}~\bibnamefont {Blazhev}}, \bibinfo {author} {\bibfnamefont
  {B.}~\bibnamefont {Bruyneel}}, \bibinfo {author} {\bibfnamefont {P.~A.}\
  \bibnamefont {Butler}}, \bibinfo {author} {\bibfnamefont {M.~P.}\
  \bibnamefont {Carpenter}}, \bibinfo {author} {\bibfnamefont {J.}~\bibnamefont
  {Cederk{\"{a}}ll}}, \bibinfo {author} {\bibfnamefont {E.}~\bibnamefont
  {Cl{\'{e}}ment}}, \bibinfo {author} {\bibfnamefont {T.~E.}\ \bibnamefont
  {Cocolios}}, \bibinfo {author} {\bibfnamefont {A.~N.}\ \bibnamefont
  {Deacon}}, \bibinfo {author} {\bibfnamefont {J.}~\bibnamefont {Diriken}},
  \bibinfo {author} {\bibfnamefont {A.}~\bibnamefont {Ekstr{\"{o}}m}}, \bibinfo
  {author} {\bibfnamefont {C.}~\bibnamefont {Fitzpatrick}}, \bibinfo {author}
  {\bibfnamefont {L.~M.}\ \bibnamefont {Fraile}}, \bibinfo {author}
  {\bibfnamefont {C.}~\bibnamefont {Fransen}}, \bibinfo {author} {\bibfnamefont
  {S.~J.}\ \bibnamefont {Freeman}}, \bibinfo {author} {\bibfnamefont {J.~E.}\
  \bibnamefont {Garc{\'{i}}a-Ramos}}, \bibinfo {author} {\bibfnamefont
  {K.}~\bibnamefont {Geibel}}, \bibinfo {author} {\bibfnamefont
  {R.}~\bibnamefont {Gernh{\"{a}}user}}, \bibinfo {author} {\bibfnamefont
  {T.}~\bibnamefont {Grahn}}, \bibinfo {author} {\bibfnamefont
  {M.}~\bibnamefont {Guttormsen}}, \bibinfo {author} {\bibfnamefont
  {B.}~\bibnamefont {Hadinia}}, \bibinfo {author} {\bibfnamefont
  {K.}~\bibnamefont {Hady{\'{n}}ska-Kl{\c{e}}k}}, \bibinfo {author}
  {\bibfnamefont {M.}~\bibnamefont {Hass}}, \bibinfo {author} {\bibfnamefont
  {P.~H.}\ \bibnamefont {Heenen}}, \bibinfo {author} {\bibfnamefont {R.~D.}\
  \bibnamefont {Herzberg}}, \bibinfo {author} {\bibfnamefont {H.}~\bibnamefont
  {Hess}}, \bibinfo {author} {\bibfnamefont {K.}~\bibnamefont {Heyde}},
  \bibinfo {author} {\bibfnamefont {M.}~\bibnamefont {Huyse}}, \bibinfo
  {author} {\bibfnamefont {O.}~\bibnamefont {Ivanov}}, \bibinfo {author}
  {\bibfnamefont {D.~G.}\ \bibnamefont {Jenkins}}, \bibinfo {author}
  {\bibfnamefont {R.}~\bibnamefont {Julin}}, \bibinfo {author} {\bibfnamefont
  {N.}~\bibnamefont {Kesteloot}}, \bibinfo {author} {\bibfnamefont
  {T.}~\bibnamefont {Kr{\"{o}}ll}}, \bibinfo {author} {\bibfnamefont
  {R.}~\bibnamefont {Kr{\"{u}}cken}}, \bibinfo {author} {\bibfnamefont {A.~C.}\
  \bibnamefont {Larsen}}, \bibinfo {author} {\bibfnamefont {R.}~\bibnamefont
  {Lutter}}, \bibinfo {author} {\bibfnamefont {P.}~\bibnamefont {Marley}},
  \bibinfo {author} {\bibfnamefont {P.~J.}\ \bibnamefont {Napiorkowski}},
  \bibinfo {author} {\bibfnamefont {R.}~\bibnamefont {Orlandi}}, \bibinfo
  {author} {\bibfnamefont {R.~D.}\ \bibnamefont {Page}}, \bibinfo {author}
  {\bibfnamefont {J.}~\bibnamefont {Pakarinen}}, \bibinfo {author}
  {\bibfnamefont {N.}~\bibnamefont {Patronis}}, \bibinfo {author}
  {\bibfnamefont {P.~J.}\ \bibnamefont {Peura}}, \bibinfo {author}
  {\bibfnamefont {E.}~\bibnamefont {Piselli}}, \bibinfo {author} {\bibfnamefont
  {L.}~\bibnamefont {Pr{\'{o}}chniak}}, \bibinfo {author} {\bibfnamefont
  {P.}~\bibnamefont {Rahkila}}, \bibinfo {author} {\bibfnamefont
  {E.}~\bibnamefont {Rapisarda}}, \bibinfo {author} {\bibfnamefont
  {P.}~\bibnamefont {Reiter}}, \bibinfo {author} {\bibfnamefont {A.~P.}\
  \bibnamefont {Robinson}}, \bibinfo {author} {\bibfnamefont {M.}~\bibnamefont
  {Scheck}}, \bibinfo {author} {\bibfnamefont {S.}~\bibnamefont {Siem}},
  \bibinfo {author} {\bibfnamefont {K.}~\bibnamefont {{Singh Chakkal}}},
  \bibinfo {author} {\bibfnamefont {J.~F.}\ \bibnamefont {Smith}}, \bibinfo
  {author} {\bibfnamefont {J.}~\bibnamefont {Srebrny}}, \bibinfo {author}
  {\bibfnamefont {I.}~\bibnamefont {Stefanescu}}, \bibinfo {author}
  {\bibfnamefont {G.~M.}\ \bibnamefont {Tveten}}, \bibinfo {author}
  {\bibfnamefont {P.}~\bibnamefont {{Van Duppen}}}, \bibinfo {author}
  {\bibfnamefont {J.}~\bibnamefont {{Van de Walle}}}, \bibinfo {author}
  {\bibfnamefont {D.}~\bibnamefont {Voulot}}, \bibinfo {author} {\bibfnamefont
  {N.}~\bibnamefont {Warr}}, \bibinfo {author} {\bibfnamefont {A.}~\bibnamefont
  {Wiens}}, \ and\ \bibinfo {author} {\bibfnamefont {J.~L.}\ \bibnamefont
  {Wood}},\ }\href {\doibase 10.1140/epja/i2019-12815-2} {\bibfield  {journal}
  {\bibinfo  {journal} {The European Physical Journal A}\ }\textbf {\bibinfo
  {volume} {55}},\ \bibinfo {pages} {130} (\bibinfo {year} {2019})}\BibitemShut
  {NoStop}%
\bibitem [{\citenamefont {{Van Duppen}}\ \emph {et~al.}(1991)\citenamefont
  {{Van Duppen}}, \citenamefont {Decrock}, \citenamefont {Dendooven},
  \citenamefont {Huyse}, \citenamefont {Reusen},\ and\ \citenamefont
  {Wauters}}]{VanDuppen1991}%
  \BibitemOpen
  \bibfield  {author} {\bibinfo {author} {\bibfnamefont {P.}~\bibnamefont {{Van
  Duppen}}}, \bibinfo {author} {\bibfnamefont {P.}~\bibnamefont {Decrock}},
  \bibinfo {author} {\bibfnamefont {P.}~\bibnamefont {Dendooven}}, \bibinfo
  {author} {\bibfnamefont {M.}~\bibnamefont {Huyse}}, \bibinfo {author}
  {\bibfnamefont {G.}~\bibnamefont {Reusen}}, \ and\ \bibinfo {author}
  {\bibfnamefont {J.}~\bibnamefont {Wauters}},\ }\href {\doibase
  10.1016/0375-9474(91)90796-9} {\bibfield  {journal} {\bibinfo  {journal}
  {Nuclear Physics A}\ }\textbf {\bibinfo {volume} {529}},\ \bibinfo {pages}
  {268} (\bibinfo {year} {1991})}\BibitemShut {NoStop}%
\bibitem [{\citenamefont {Kreiner}\ \emph {et~al.}(1981)\citenamefont
  {Kreiner}, \citenamefont {Baktash}, \citenamefont {{Garcia Bermudez}},\ and\
  \citenamefont {Mariscotti}}]{Kreiner1981}%
  \BibitemOpen
  \bibfield  {author} {\bibinfo {author} {\bibfnamefont {A.~J.}\ \bibnamefont
  {Kreiner}}, \bibinfo {author} {\bibfnamefont {C.}~\bibnamefont {Baktash}},
  \bibinfo {author} {\bibfnamefont {G.}~\bibnamefont {{Garcia Bermudez}}}, \
  and\ \bibinfo {author} {\bibfnamefont {M.~A.~J.}\ \bibnamefont
  {Mariscotti}},\ }\href {\doibase 10.1103/PhysRevLett.47.1709} {\bibfield
  {journal} {\bibinfo  {journal} {Physical Review Letters}\ }\textbf {\bibinfo
  {volume} {47}},\ \bibinfo {pages} {1709} (\bibinfo {year}
  {1981})}\BibitemShut {NoStop}%
\bibitem [{\citenamefont {{M. {S}tryjczyk \textit{et
  al.}}}()}]{StryjczykHgbeta}%
  \BibitemOpen
  \bibfield  {author} {\bibinfo {author} {\bibnamefont {{M. {S}tryjczyk
  \textit{et al.}}}},\ }\href@noop {} {\enquote {\bibinfo {title}
  {{$\beta$-decay studies of $^{182,184,186}${T}l}},}\ }\bibinfo {note}
  {(unpublished)}\BibitemShut {NoStop}%
\bibitem [{\citenamefont {Fedosseev}\ \emph {et~al.}(2017)\citenamefont
  {Fedosseev}, \citenamefont {Chrysalidis}, \citenamefont {Goodacre},
  \citenamefont {Marsh}, \citenamefont {Rothe}, \citenamefont {Seiffert},\ and\
  \citenamefont {Wendt}}]{Fedosseev2017}%
  \BibitemOpen
  \bibfield  {author} {\bibinfo {author} {\bibfnamefont {V.}~\bibnamefont
  {Fedosseev}}, \bibinfo {author} {\bibfnamefont {K.}~\bibnamefont
  {Chrysalidis}}, \bibinfo {author} {\bibfnamefont {T.~D.}\ \bibnamefont
  {Goodacre}}, \bibinfo {author} {\bibfnamefont {B.}~\bibnamefont {Marsh}},
  \bibinfo {author} {\bibfnamefont {S.}~\bibnamefont {Rothe}}, \bibinfo
  {author} {\bibfnamefont {C.}~\bibnamefont {Seiffert}}, \ and\ \bibinfo
  {author} {\bibfnamefont {K.}~\bibnamefont {Wendt}},\ }\href {\doibase
  10.1088/1361-6471/aa78e0} {\bibfield  {journal} {\bibinfo  {journal} {Journal
  of Physics G: Nuclear and Particle Physics}\ }\textbf {\bibinfo {volume}
  {44}},\ \bibinfo {pages} {084006} (\bibinfo {year} {2017})}\BibitemShut
  {NoStop}%
\bibitem [{\citenamefont {Catherall}\ \emph {et~al.}(2017)\citenamefont
  {Catherall}, \citenamefont {Andreazza}, \citenamefont {Breitenfeldt},
  \citenamefont {Dorsival}, \citenamefont {Focker}, \citenamefont {Gharsa},
  \citenamefont {{T J}}, \citenamefont {Grenard}, \citenamefont {Locci},
  \citenamefont {Martins}, \citenamefont {Marzari}, \citenamefont {Schipper},
  \citenamefont {Shornikov},\ and\ \citenamefont {Stora}}]{Catherall2017}%
  \BibitemOpen
  \bibfield  {author} {\bibinfo {author} {\bibfnamefont {R.}~\bibnamefont
  {Catherall}}, \bibinfo {author} {\bibfnamefont {W.}~\bibnamefont
  {Andreazza}}, \bibinfo {author} {\bibfnamefont {M.}~\bibnamefont
  {Breitenfeldt}}, \bibinfo {author} {\bibfnamefont {A.}~\bibnamefont
  {Dorsival}}, \bibinfo {author} {\bibfnamefont {G.~J.}\ \bibnamefont
  {Focker}}, \bibinfo {author} {\bibfnamefont {T.~P.}\ \bibnamefont {Gharsa}},
  \bibinfo {author} {\bibfnamefont {G.}~\bibnamefont {{T J}}}, \bibinfo
  {author} {\bibfnamefont {J.-L.}\ \bibnamefont {Grenard}}, \bibinfo {author}
  {\bibfnamefont {F.}~\bibnamefont {Locci}}, \bibinfo {author} {\bibfnamefont
  {P.}~\bibnamefont {Martins}}, \bibinfo {author} {\bibfnamefont
  {S.}~\bibnamefont {Marzari}}, \bibinfo {author} {\bibfnamefont
  {J.}~\bibnamefont {Schipper}}, \bibinfo {author} {\bibfnamefont
  {A.}~\bibnamefont {Shornikov}}, \ and\ \bibinfo {author} {\bibfnamefont
  {T.}~\bibnamefont {Stora}},\ }\href {\doibase 10.1088/1361-6471/aa7eba}
  {\bibfield  {journal} {\bibinfo  {journal} {Journal of Physics G: Nuclear and
  Particle Physics}\ }\textbf {\bibinfo {volume} {44}},\ \bibinfo {pages}
  {094002} (\bibinfo {year} {2017})}\BibitemShut {NoStop}%
\bibitem [{IDS()}]{IDS}%
  \BibitemOpen
  \href@noop {} {}\bibinfo {note}
  {\url{http://isolde-ids.web.cern.ch/isolde-ids/}}\BibitemShut {NoStop}%
\bibitem [{\citenamefont {Papadakis}\ \emph {et~al.}(2018)\citenamefont
  {Papadakis}, \citenamefont {Cox}, \citenamefont {O'Neill}, \citenamefont
  {Borge}, \citenamefont {Butler}, \citenamefont {Gaffney}, \citenamefont
  {Greenlees}, \citenamefont {Herzberg}, \citenamefont {Illana}, \citenamefont
  {Joss}, \citenamefont {Konki}, \citenamefont {Kr{\"{o}}ll}, \citenamefont
  {Ojala}, \citenamefont {Page}, \citenamefont {Rahkila}, \citenamefont
  {Ranttila}, \citenamefont {Thornhill}, \citenamefont {Tuunanen},
  \citenamefont {{Van Duppen}}, \citenamefont {Warr},\ and\ \citenamefont
  {Pakarinen}}]{Papadakis2018}%
  \BibitemOpen
  \bibfield  {author} {\bibinfo {author} {\bibfnamefont {P.}~\bibnamefont
  {Papadakis}}, \bibinfo {author} {\bibfnamefont {D.~M.}\ \bibnamefont {Cox}},
  \bibinfo {author} {\bibfnamefont {G.~G.}\ \bibnamefont {O'Neill}}, \bibinfo
  {author} {\bibfnamefont {M.~J.~G.}\ \bibnamefont {Borge}}, \bibinfo {author}
  {\bibfnamefont {P.~A.}\ \bibnamefont {Butler}}, \bibinfo {author}
  {\bibfnamefont {L.~P.}\ \bibnamefont {Gaffney}}, \bibinfo {author}
  {\bibfnamefont {P.~T.}\ \bibnamefont {Greenlees}}, \bibinfo {author}
  {\bibfnamefont {R.~D.}\ \bibnamefont {Herzberg}}, \bibinfo {author}
  {\bibfnamefont {A.}~\bibnamefont {Illana}}, \bibinfo {author} {\bibfnamefont
  {D.~T.}\ \bibnamefont {Joss}}, \bibinfo {author} {\bibfnamefont
  {J.}~\bibnamefont {Konki}}, \bibinfo {author} {\bibfnamefont
  {T.}~\bibnamefont {Kr{\"{o}}ll}}, \bibinfo {author} {\bibfnamefont
  {J.}~\bibnamefont {Ojala}}, \bibinfo {author} {\bibfnamefont {R.~D.}\
  \bibnamefont {Page}}, \bibinfo {author} {\bibfnamefont {P.}~\bibnamefont
  {Rahkila}}, \bibinfo {author} {\bibfnamefont {K.}~\bibnamefont {Ranttila}},
  \bibinfo {author} {\bibfnamefont {J.}~\bibnamefont {Thornhill}}, \bibinfo
  {author} {\bibfnamefont {J.}~\bibnamefont {Tuunanen}}, \bibinfo {author}
  {\bibfnamefont {P.}~\bibnamefont {{Van Duppen}}}, \bibinfo {author}
  {\bibfnamefont {N.}~\bibnamefont {Warr}}, \ and\ \bibinfo {author}
  {\bibfnamefont {J.}~\bibnamefont {Pakarinen}},\ }\href {\doibase
  10.1140/epja/i2018-12474-9} {\bibfield  {journal} {\bibinfo  {journal} {The
  European Physical Journal A}\ }\textbf {\bibinfo {volume} {54}},\ \bibinfo
  {pages} {42} (\bibinfo {year} {2018})}\BibitemShut {NoStop}%
\bibitem [{Nut()}]{Nutaq}%
  \BibitemOpen
  \href@noop {} {}\bibinfo {note} {\url{http://www.nutaq.com}}\BibitemShut
  {NoStop}%
\bibitem [{\citenamefont {Ijaz}\ \emph {et~al.}(1977)\citenamefont {Ijaz},
  \citenamefont {Bingham}, \citenamefont {Carter}, \citenamefont {Robinson},\
  and\ \citenamefont {Toth}}]{Ijaz1977}%
  \BibitemOpen
  \bibfield  {author} {\bibinfo {author} {\bibfnamefont {M.~A.}\ \bibnamefont
  {Ijaz}}, \bibinfo {author} {\bibfnamefont {C.~R.}\ \bibnamefont {Bingham}},
  \bibinfo {author} {\bibfnamefont {H.~K.}\ \bibnamefont {Carter}}, \bibinfo
  {author} {\bibfnamefont {E.~L.}\ \bibnamefont {Robinson}}, \ and\ \bibinfo
  {author} {\bibfnamefont {K.~S.}\ \bibnamefont {Toth}},\ }\href {\doibase
  10.1103/PhysRevC.15.2251} {\bibfield  {journal} {\bibinfo  {journal}
  {Physical Review C}\ }\textbf {\bibinfo {volume} {15}},\ \bibinfo {pages}
  {2251} (\bibinfo {year} {1977})}\BibitemShut {NoStop}%
\bibitem [{\citenamefont {Hansen}\ \emph {et~al.}(1970)\citenamefont {Hansen},
  \citenamefont {Nielsen}, \citenamefont {Wilsky}, \citenamefont {Alpsten},
  \citenamefont {Finger}, \citenamefont {Lindahl}, \citenamefont {Naumann},\
  and\ \citenamefont {Nielsen}}]{Hansen1970}%
  \BibitemOpen
  \bibfield  {author} {\bibinfo {author} {\bibfnamefont {P.}~\bibnamefont
  {Hansen}}, \bibinfo {author} {\bibfnamefont {H.}~\bibnamefont {Nielsen}},
  \bibinfo {author} {\bibfnamefont {K.}~\bibnamefont {Wilsky}}, \bibinfo
  {author} {\bibfnamefont {M.}~\bibnamefont {Alpsten}}, \bibinfo {author}
  {\bibfnamefont {M.}~\bibnamefont {Finger}}, \bibinfo {author} {\bibfnamefont
  {A.}~\bibnamefont {Lindahl}}, \bibinfo {author} {\bibfnamefont
  {R.}~\bibnamefont {Naumann}}, \ and\ \bibinfo {author} {\bibfnamefont
  {O.}~\bibnamefont {Nielsen}},\ }\href {\doibase 10.1016/0375-9474(70)90622-6}
  {\bibfield  {journal} {\bibinfo  {journal} {Nuclear Physics A}\ }\textbf
  {\bibinfo {volume} {148}},\ \bibinfo {pages} {249} (\bibinfo {year}
  {1970})}\BibitemShut {NoStop}%
\bibitem [{\citenamefont {Ibrahim}\ \emph {et~al.}(2001)\citenamefont
  {Ibrahim}, \citenamefont {Genevey}, \citenamefont {Cottereau}, \citenamefont
  {Gizon}, \citenamefont {Knipper}, \citenamefont {{Le Blanc}}, \citenamefont
  {Marguier}, \citenamefont {Obert}, \citenamefont {Oms}, \citenamefont
  {Putaux}, \citenamefont {Roussi{\`{e}}re}, \citenamefont {Sauvage},\ and\
  \citenamefont {Wojtasiewicz}}]{Ibrahim2001}%
  \BibitemOpen
  \bibfield  {author} {\bibinfo {author} {\bibfnamefont {F.}~\bibnamefont
  {Ibrahim}}, \bibinfo {author} {\bibfnamefont {J.}~\bibnamefont {Genevey}},
  \bibinfo {author} {\bibfnamefont {E.}~\bibnamefont {Cottereau}}, \bibinfo
  {author} {\bibfnamefont {A.}~\bibnamefont {Gizon}}, \bibinfo {author}
  {\bibfnamefont {A.}~\bibnamefont {Knipper}}, \bibinfo {author} {\bibfnamefont
  {F.}~\bibnamefont {{Le Blanc}}}, \bibinfo {author} {\bibfnamefont
  {G.}~\bibnamefont {Marguier}}, \bibinfo {author} {\bibfnamefont
  {J.}~\bibnamefont {Obert}}, \bibinfo {author} {\bibfnamefont
  {J.}~\bibnamefont {Oms}}, \bibinfo {author} {\bibfnamefont {J.}~\bibnamefont
  {Putaux}}, \bibinfo {author} {\bibfnamefont {B.}~\bibnamefont
  {Roussi{\`{e}}re}}, \bibinfo {author} {\bibfnamefont {J.}~\bibnamefont
  {Sauvage}}, \ and\ \bibinfo {author} {\bibfnamefont {A.}~\bibnamefont
  {Wojtasiewicz}},\ }\href {\doibase 10.1007/s100500170125} {\bibfield
  {journal} {\bibinfo  {journal} {The European Physical Journal A}\ }\textbf
  {\bibinfo {volume} {10}},\ \bibinfo {pages} {139} (\bibinfo {year}
  {2001})}\BibitemShut {NoStop}%
\bibitem [{\citenamefont {Harding}\ \emph {et~al.}(2020)\citenamefont
  {Harding}, \citenamefont {Andreyev}, \citenamefont {Barzakh}, \citenamefont
  {Atanasov}, \citenamefont {Cubiss}, \citenamefont {{Van Duppen}},
  \citenamefont {{Al Monthery}}, \citenamefont {Althubiti}, \citenamefont
  {Andel}, \citenamefont {Antalic}, \citenamefont {Blaum}, \citenamefont
  {Cocolios}, \citenamefont {{Day Goodacre}}, \citenamefont {{De Roubin}},
  \citenamefont {Farooq-Smith}, \citenamefont {Fedorov}, \citenamefont
  {Fedosseev}, \citenamefont {Fink}, \citenamefont {Gaffney}, \citenamefont
  {Ghys}, \citenamefont {Joss}, \citenamefont {Herfurth}, \citenamefont
  {Huyse}, \citenamefont {Imai}, \citenamefont {Kreim}, \citenamefont {Lunney},
  \citenamefont {Lynch}, \citenamefont {Manea}, \citenamefont {Marsh},
  \citenamefont {{Martinez Palenzuela}}, \citenamefont {Molkanov},
  \citenamefont {Neidherr}, \citenamefont {Page}, \citenamefont {Pastore},
  \citenamefont {Rosenbusch}, \citenamefont {Rossel}, \citenamefont {Rothe},
  \citenamefont {Schweikhard}, \citenamefont {Seliverstov}, \citenamefont
  {Sels}, \citenamefont {{Van Beveren}}, \citenamefont {Verstraelen},
  \citenamefont {Welker}, \citenamefont {Wienholtz}, \citenamefont {Wolf},\
  and\ \citenamefont {Zuber}}]{Harding2020}%
  \BibitemOpen
  \bibfield  {author} {\bibinfo {author} {\bibfnamefont {R.~D.}\ \bibnamefont
  {Harding}}, \bibinfo {author} {\bibfnamefont {A.~N.}\ \bibnamefont
  {Andreyev}}, \bibinfo {author} {\bibfnamefont {A.~E.}\ \bibnamefont
  {Barzakh}}, \bibinfo {author} {\bibfnamefont {D.}~\bibnamefont {Atanasov}},
  \bibinfo {author} {\bibfnamefont {J.~G.}\ \bibnamefont {Cubiss}}, \bibinfo
  {author} {\bibfnamefont {P.}~\bibnamefont {{Van Duppen}}}, \bibinfo {author}
  {\bibfnamefont {M.}~\bibnamefont {{Al Monthery}}}, \bibinfo {author}
  {\bibfnamefont {N.~A.}\ \bibnamefont {Althubiti}}, \bibinfo {author}
  {\bibfnamefont {B.}~\bibnamefont {Andel}}, \bibinfo {author} {\bibfnamefont
  {S.}~\bibnamefont {Antalic}}, \bibinfo {author} {\bibfnamefont
  {K.}~\bibnamefont {Blaum}}, \bibinfo {author} {\bibfnamefont {T.~E.}\
  \bibnamefont {Cocolios}}, \bibinfo {author} {\bibfnamefont {T.}~\bibnamefont
  {{Day Goodacre}}}, \bibinfo {author} {\bibfnamefont {A.}~\bibnamefont {{De
  Roubin}}}, \bibinfo {author} {\bibfnamefont {G.~J.}\ \bibnamefont
  {Farooq-Smith}}, \bibinfo {author} {\bibfnamefont {D.~V.}\ \bibnamefont
  {Fedorov}}, \bibinfo {author} {\bibfnamefont {V.~N.}\ \bibnamefont
  {Fedosseev}}, \bibinfo {author} {\bibfnamefont {D.~A.}\ \bibnamefont {Fink}},
  \bibinfo {author} {\bibfnamefont {L.~P.}\ \bibnamefont {Gaffney}}, \bibinfo
  {author} {\bibfnamefont {L.}~\bibnamefont {Ghys}}, \bibinfo {author}
  {\bibfnamefont {D.~T.}\ \bibnamefont {Joss}}, \bibinfo {author}
  {\bibfnamefont {F.}~\bibnamefont {Herfurth}}, \bibinfo {author}
  {\bibfnamefont {M.}~\bibnamefont {Huyse}}, \bibinfo {author} {\bibfnamefont
  {N.}~\bibnamefont {Imai}}, \bibinfo {author} {\bibfnamefont {S.}~\bibnamefont
  {Kreim}}, \bibinfo {author} {\bibfnamefont {D.}~\bibnamefont {Lunney}},
  \bibinfo {author} {\bibfnamefont {K.~M.}\ \bibnamefont {Lynch}}, \bibinfo
  {author} {\bibfnamefont {V.}~\bibnamefont {Manea}}, \bibinfo {author}
  {\bibfnamefont {B.~A.}\ \bibnamefont {Marsh}}, \bibinfo {author}
  {\bibfnamefont {Y.}~\bibnamefont {{Martinez Palenzuela}}}, \bibinfo {author}
  {\bibfnamefont {P.~L.}\ \bibnamefont {Molkanov}}, \bibinfo {author}
  {\bibfnamefont {D.}~\bibnamefont {Neidherr}}, \bibinfo {author}
  {\bibfnamefont {R.~D.}\ \bibnamefont {Page}}, \bibinfo {author}
  {\bibfnamefont {A.}~\bibnamefont {Pastore}}, \bibinfo {author} {\bibfnamefont
  {M.}~\bibnamefont {Rosenbusch}}, \bibinfo {author} {\bibfnamefont {R.~E.}\
  \bibnamefont {Rossel}}, \bibinfo {author} {\bibfnamefont {S.}~\bibnamefont
  {Rothe}}, \bibinfo {author} {\bibfnamefont {L.}~\bibnamefont {Schweikhard}},
  \bibinfo {author} {\bibfnamefont {M.~D.}\ \bibnamefont {Seliverstov}},
  \bibinfo {author} {\bibfnamefont {S.}~\bibnamefont {Sels}}, \bibinfo {author}
  {\bibfnamefont {C.}~\bibnamefont {{Van Beveren}}}, \bibinfo {author}
  {\bibfnamefont {E.}~\bibnamefont {Verstraelen}}, \bibinfo {author}
  {\bibfnamefont {A.}~\bibnamefont {Welker}}, \bibinfo {author} {\bibfnamefont
  {F.}~\bibnamefont {Wienholtz}}, \bibinfo {author} {\bibfnamefont {R.~N.}\
  \bibnamefont {Wolf}}, \ and\ \bibinfo {author} {\bibfnamefont
  {K.}~\bibnamefont {Zuber}},\ }\href {\doibase 10.1103/PhysRevC.102.024312}
  {\bibfield  {journal} {\bibinfo  {journal} {Physical Review C}\ }\textbf
  {\bibinfo {volume} {102}},\ \bibinfo {pages} {024312} (\bibinfo {year}
  {2020})}\BibitemShut {NoStop}%
\bibitem [{\citenamefont {Kib{\'{e}}di}\ \emph {et~al.}(2008)\citenamefont
  {Kib{\'{e}}di}, \citenamefont {Burrows}, \citenamefont {Trzhaskovskaya},
  \citenamefont {Davidson},\ and\ \citenamefont {Nestor}}]{Kibedi2008}%
  \BibitemOpen
  \bibfield  {author} {\bibinfo {author} {\bibfnamefont {T.}~\bibnamefont
  {Kib{\'{e}}di}}, \bibinfo {author} {\bibfnamefont {T.}~\bibnamefont
  {Burrows}}, \bibinfo {author} {\bibfnamefont {M.}~\bibnamefont
  {Trzhaskovskaya}}, \bibinfo {author} {\bibfnamefont {P.}~\bibnamefont
  {Davidson}}, \ and\ \bibinfo {author} {\bibfnamefont {C.}~\bibnamefont
  {Nestor}},\ }\href {\doibase 10.1016/j.nima.2008.02.051} {\bibfield
  {journal} {\bibinfo  {journal} {Nuclear Instruments and Methods in Physics
  Research Section A: Accelerators, Spectrometers, Detectors and Associated
  Equipment}\ }\textbf {\bibinfo {volume} {589}},\ \bibinfo {pages} {202}
  (\bibinfo {year} {2008})}\BibitemShut {NoStop}%
\bibitem [{\citenamefont {Zhang}\ \emph {et~al.}(2002)\citenamefont {Zhang},
  \citenamefont {Xu}, \citenamefont {He}, \citenamefont {Liu}, \citenamefont
  {Zhou}, \citenamefont {Gan}, \citenamefont {Hayakawa}, \citenamefont
  {Oshima}, \citenamefont {Toh}, \citenamefont {Shizuma}, \citenamefont
  {Katakura}, \citenamefont {Hatsukawa}, \citenamefont {Matsuda}, \citenamefont
  {Kusakari}, \citenamefont {Sugawara}, \citenamefont {Furuno}, \citenamefont
  {Komatsubara}, \citenamefont {Une}, \citenamefont {Wen},\ and\ \citenamefont
  {Wang}}]{Zhang2002}%
  \BibitemOpen
  \bibfield  {author} {\bibinfo {author} {\bibfnamefont {Y.}~\bibnamefont
  {Zhang}}, \bibinfo {author} {\bibfnamefont {F.}~\bibnamefont {Xu}}, \bibinfo
  {author} {\bibfnamefont {J.}~\bibnamefont {He}}, \bibinfo {author}
  {\bibfnamefont {Z.}~\bibnamefont {Liu}}, \bibinfo {author} {\bibfnamefont
  {X.}~\bibnamefont {Zhou}}, \bibinfo {author} {\bibfnamefont {Z.}~\bibnamefont
  {Gan}}, \bibinfo {author} {\bibfnamefont {T.}~\bibnamefont {Hayakawa}},
  \bibinfo {author} {\bibfnamefont {M.}~\bibnamefont {Oshima}}, \bibinfo
  {author} {\bibfnamefont {T.}~\bibnamefont {Toh}}, \bibinfo {author}
  {\bibfnamefont {T.}~\bibnamefont {Shizuma}}, \bibinfo {author} {\bibfnamefont
  {J.}~\bibnamefont {Katakura}}, \bibinfo {author} {\bibfnamefont
  {Y.}~\bibnamefont {Hatsukawa}}, \bibinfo {author} {\bibfnamefont
  {M.}~\bibnamefont {Matsuda}}, \bibinfo {author} {\bibfnamefont
  {H.}~\bibnamefont {Kusakari}}, \bibinfo {author} {\bibfnamefont
  {M.}~\bibnamefont {Sugawara}}, \bibinfo {author} {\bibfnamefont
  {K.}~\bibnamefont {Furuno}}, \bibinfo {author} {\bibfnamefont
  {T.}~\bibnamefont {Komatsubara}}, \bibinfo {author} {\bibfnamefont
  {T.}~\bibnamefont {Une}}, \bibinfo {author} {\bibfnamefont {S.}~\bibnamefont
  {Wen}}, \ and\ \bibinfo {author} {\bibfnamefont {Z.}~\bibnamefont {Wang}},\
  }\href {\doibase 10.1140/epja/i2002-10041-9} {\bibfield  {journal} {\bibinfo
  {journal} {The European Physical Journal A}\ }\textbf {\bibinfo {volume}
  {14}},\ \bibinfo {pages} {271} (\bibinfo {year} {2002})}\BibitemShut
  {NoStop}%
\bibitem [{\citenamefont {Wang}\ \emph {et~al.}(2017)\citenamefont {Wang},
  \citenamefont {Audi}, \citenamefont {Kondev}, \citenamefont {Huang},
  \citenamefont {Naimi},\ and\ \citenamefont {Xu}}]{Wang2017}%
  \BibitemOpen
  \bibfield  {author} {\bibinfo {author} {\bibfnamefont {M.}~\bibnamefont
  {Wang}}, \bibinfo {author} {\bibfnamefont {G.}~\bibnamefont {Audi}}, \bibinfo
  {author} {\bibfnamefont {F.~G.}\ \bibnamefont {Kondev}}, \bibinfo {author}
  {\bibfnamefont {W.}~\bibnamefont {Huang}}, \bibinfo {author} {\bibfnamefont
  {S.}~\bibnamefont {Naimi}}, \ and\ \bibinfo {author} {\bibfnamefont
  {X.}~\bibnamefont {Xu}},\ }\href {\doibase 10.1088/1674-1137/41/3/030003}
  {\bibfield  {journal} {\bibinfo  {journal} {Chinese Physics C}\ }\textbf
  {\bibinfo {volume} {41}},\ \bibinfo {pages} {030003} (\bibinfo {year}
  {2017})}\BibitemShut {NoStop}%
\bibitem [{\citenamefont {Pfeiffer}\ \emph {et~al.}(2014)\citenamefont
  {Pfeiffer}, \citenamefont {Venkataramaniah}, \citenamefont {Czok},\ and\
  \citenamefont {Scheidenberger}}]{Pfeiffer2014}%
  \BibitemOpen
  \bibfield  {author} {\bibinfo {author} {\bibfnamefont {B.}~\bibnamefont
  {Pfeiffer}}, \bibinfo {author} {\bibfnamefont {K.}~\bibnamefont
  {Venkataramaniah}}, \bibinfo {author} {\bibfnamefont {U.}~\bibnamefont
  {Czok}}, \ and\ \bibinfo {author} {\bibfnamefont {C.}~\bibnamefont
  {Scheidenberger}},\ }\href {\doibase 10.1016/j.adt.2013.06.002} {\bibfield
  {journal} {\bibinfo  {journal} {Atomic Data and Nuclear Data Tables}\
  }\textbf {\bibinfo {volume} {100}},\ \bibinfo {pages} {403} (\bibinfo {year}
  {2014})}\BibitemShut {NoStop}%
\bibitem [{\citenamefont {Weber}\ \emph {et~al.}(2008)\citenamefont {Weber},
  \citenamefont {Audi}, \citenamefont {Beck}, \citenamefont {Blaum},
  \citenamefont {Bollen}, \citenamefont {Herfurth}, \citenamefont
  {Kellerbauer}, \citenamefont {Kluge}, \citenamefont {Lunney},\ and\
  \citenamefont {Schwarz}}]{Weber2008}%
  \BibitemOpen
  \bibfield  {author} {\bibinfo {author} {\bibfnamefont {C.}~\bibnamefont
  {Weber}}, \bibinfo {author} {\bibfnamefont {G.}~\bibnamefont {Audi}},
  \bibinfo {author} {\bibfnamefont {D.}~\bibnamefont {Beck}}, \bibinfo {author}
  {\bibfnamefont {K.}~\bibnamefont {Blaum}}, \bibinfo {author} {\bibfnamefont
  {G.}~\bibnamefont {Bollen}}, \bibinfo {author} {\bibfnamefont
  {F.}~\bibnamefont {Herfurth}}, \bibinfo {author} {\bibfnamefont
  {A.}~\bibnamefont {Kellerbauer}}, \bibinfo {author} {\bibfnamefont {H.-J.}\
  \bibnamefont {Kluge}}, \bibinfo {author} {\bibfnamefont {D.}~\bibnamefont
  {Lunney}}, \ and\ \bibinfo {author} {\bibfnamefont {S.}~\bibnamefont
  {Schwarz}},\ }\href {\doibase 10.1016/j.nuclphysa.2007.12.014} {\bibfield
  {journal} {\bibinfo  {journal} {Nuclear Physics A}\ }\textbf {\bibinfo
  {volume} {803}},\ \bibinfo {pages} {1} (\bibinfo {year} {2008})}\BibitemShut
  {NoStop}%
\bibitem [{\citenamefont {Baglin}(2003)}]{Baglin2003}%
  \BibitemOpen
  \bibfield  {author} {\bibinfo {author} {\bibfnamefont {C.~M.}\ \bibnamefont
  {Baglin}},\ }\href {\doibase 10.1006/ndsh.2003.0007} {\bibfield  {journal}
  {\bibinfo  {journal} {Nuclear Data Sheets}\ }\textbf {\bibinfo {volume}
  {99}},\ \bibinfo {pages} {1} (\bibinfo {year} {2003})}\BibitemShut {NoStop}%
\bibitem [{\citenamefont {Andel}(2016)}]{BorisThesis}%
  \BibitemOpen
  \bibfield  {author} {\bibinfo {author} {\bibfnamefont {B.}~\bibnamefont
  {Andel}},\ }\emph {\bibinfo {title} {{S}tudy of neutron deficient polonium
  isotopes}},\ \href {https://repository.gsi.de/record/206532} {Ph.D. thesis},\
  \bibinfo  {school} {Comenius University in Bratislava} (\bibinfo {year}
  {2016})\BibitemShut {NoStop}%
\bibitem [{\citenamefont {Hamilton}\ \emph {et~al.}(1975)\citenamefont
  {Hamilton}, \citenamefont {Ramayya}, \citenamefont {Bosworth}, \citenamefont
  {Lourens}, \citenamefont {Cole}, \citenamefont {{Van Nooijen}}, \citenamefont
  {Garcia-Bermudez}, \citenamefont {Martin}, \citenamefont {Rao}, \citenamefont
  {Kawakami}, \citenamefont {Riedinger}, \citenamefont {Bingham}, \citenamefont
  {Turner}, \citenamefont {Zganjar}, \citenamefont {Spejewski}, \citenamefont
  {Carter}, \citenamefont {Mlekodaj}, \citenamefont {Schmidt-Ott},
  \citenamefont {Baker}, \citenamefont {Fink}, \citenamefont {Gowdy},
  \citenamefont {Wood}, \citenamefont {Xenoulis}, \citenamefont {Kern},
  \citenamefont {Hofstetter}, \citenamefont {Weil}, \citenamefont {Toth},
  \citenamefont {Ijaz},\ and\ \citenamefont {Faftry}}]{Hamilton1975}%
  \BibitemOpen
  \bibfield  {author} {\bibinfo {author} {\bibfnamefont {J.~H.}\ \bibnamefont
  {Hamilton}}, \bibinfo {author} {\bibfnamefont {A.~V.}\ \bibnamefont
  {Ramayya}}, \bibinfo {author} {\bibfnamefont {E.~L.}\ \bibnamefont
  {Bosworth}}, \bibinfo {author} {\bibfnamefont {W.}~\bibnamefont {Lourens}},
  \bibinfo {author} {\bibfnamefont {J.~D.}\ \bibnamefont {Cole}}, \bibinfo
  {author} {\bibfnamefont {B.}~\bibnamefont {{Van Nooijen}}}, \bibinfo {author}
  {\bibfnamefont {G.}~\bibnamefont {Garcia-Bermudez}}, \bibinfo {author}
  {\bibfnamefont {B.}~\bibnamefont {Martin}}, \bibinfo {author} {\bibfnamefont
  {B.~N.~S.}\ \bibnamefont {Rao}}, \bibinfo {author} {\bibfnamefont
  {H.}~\bibnamefont {Kawakami}}, \bibinfo {author} {\bibfnamefont {L.~L.}\
  \bibnamefont {Riedinger}}, \bibinfo {author} {\bibfnamefont {C.~R.}\
  \bibnamefont {Bingham}}, \bibinfo {author} {\bibfnamefont {F.}~\bibnamefont
  {Turner}}, \bibinfo {author} {\bibfnamefont {E.~F.}\ \bibnamefont {Zganjar}},
  \bibinfo {author} {\bibfnamefont {E.~H.}\ \bibnamefont {Spejewski}}, \bibinfo
  {author} {\bibfnamefont {H.~K.}\ \bibnamefont {Carter}}, \bibinfo {author}
  {\bibfnamefont {R.~L.}\ \bibnamefont {Mlekodaj}}, \bibinfo {author}
  {\bibfnamefont {W.~D.}\ \bibnamefont {Schmidt-Ott}}, \bibinfo {author}
  {\bibfnamefont {K.~R.}\ \bibnamefont {Baker}}, \bibinfo {author}
  {\bibfnamefont {R.~W.}\ \bibnamefont {Fink}}, \bibinfo {author}
  {\bibfnamefont {G.~M.}\ \bibnamefont {Gowdy}}, \bibinfo {author}
  {\bibfnamefont {J.~L.}\ \bibnamefont {Wood}}, \bibinfo {author}
  {\bibfnamefont {A.}~\bibnamefont {Xenoulis}}, \bibinfo {author}
  {\bibfnamefont {B.~D.}\ \bibnamefont {Kern}}, \bibinfo {author}
  {\bibfnamefont {K.~J.}\ \bibnamefont {Hofstetter}}, \bibinfo {author}
  {\bibfnamefont {J.~L.}\ \bibnamefont {Weil}}, \bibinfo {author}
  {\bibfnamefont {K.~S.}\ \bibnamefont {Toth}}, \bibinfo {author}
  {\bibfnamefont {M.~A.}\ \bibnamefont {Ijaz}}, \ and\ \bibinfo {author}
  {\bibfnamefont {K.~F.~R.}\ \bibnamefont {Faftry}},\ }\href {\doibase
  10.1103/PhysRevLett.35.562} {\bibfield  {journal} {\bibinfo  {journal}
  {Physical Review Letters}\ }\textbf {\bibinfo {volume} {35}},\ \bibinfo
  {pages} {562} (\bibinfo {year} {1975})}\BibitemShut {NoStop}%
\bibitem [{\citenamefont {Cole}\ \emph {et~al.}(1977)\citenamefont {Cole},
  \citenamefont {Ramayya}, \citenamefont {Hamilton}, \citenamefont {Kawakami},
  \citenamefont {van Nooijen}, \citenamefont {Nettles}, \citenamefont
  {Riedinger}, \citenamefont {Turner}, \citenamefont {Bingham}, \citenamefont
  {Carter}, \citenamefont {Spejewski}, \citenamefont {Mlekodaj}, \citenamefont
  {Schmidt-Ott}, \citenamefont {Zganjar}, \citenamefont {Sastry}, \citenamefont
  {Avignone}, \citenamefont {Toth},\ and\ \citenamefont {Ijaz}}]{Cole1977}%
  \BibitemOpen
  \bibfield  {author} {\bibinfo {author} {\bibfnamefont {J.~D.}\ \bibnamefont
  {Cole}}, \bibinfo {author} {\bibfnamefont {A.~V.}\ \bibnamefont {Ramayya}},
  \bibinfo {author} {\bibfnamefont {J.~H.}\ \bibnamefont {Hamilton}}, \bibinfo
  {author} {\bibfnamefont {H.}~\bibnamefont {Kawakami}}, \bibinfo {author}
  {\bibfnamefont {B.}~\bibnamefont {van Nooijen}}, \bibinfo {author}
  {\bibfnamefont {W.~G.}\ \bibnamefont {Nettles}}, \bibinfo {author}
  {\bibfnamefont {L.~L.}\ \bibnamefont {Riedinger}}, \bibinfo {author}
  {\bibfnamefont {F.~E.}\ \bibnamefont {Turner}}, \bibinfo {author}
  {\bibfnamefont {C.~R.}\ \bibnamefont {Bingham}}, \bibinfo {author}
  {\bibfnamefont {H.~K.}\ \bibnamefont {Carter}}, \bibinfo {author}
  {\bibfnamefont {E.~H.}\ \bibnamefont {Spejewski}}, \bibinfo {author}
  {\bibfnamefont {R.~L.}\ \bibnamefont {Mlekodaj}}, \bibinfo {author}
  {\bibfnamefont {W.~D.}\ \bibnamefont {Schmidt-Ott}}, \bibinfo {author}
  {\bibfnamefont {E.~F.}\ \bibnamefont {Zganjar}}, \bibinfo {author}
  {\bibfnamefont {K.~S.~R.}\ \bibnamefont {Sastry}}, \bibinfo {author}
  {\bibfnamefont {F.~T.}\ \bibnamefont {Avignone}}, \bibinfo {author}
  {\bibfnamefont {K.~S.}\ \bibnamefont {Toth}}, \ and\ \bibinfo {author}
  {\bibfnamefont {M.~A.}\ \bibnamefont {Ijaz}},\ }\href {\doibase
  10.1103/PhysRevC.16.2010} {\bibfield  {journal} {\bibinfo  {journal}
  {Physical Review C}\ }\textbf {\bibinfo {volume} {16}},\ \bibinfo {pages}
  {2010} (\bibinfo {year} {1977})}\BibitemShut {NoStop}%
\bibitem [{\citenamefont {B{\'{e}}raud}\ \emph {et~al.}(1977)\citenamefont
  {B{\'{e}}raud}, \citenamefont {Meyer}, \citenamefont {Desthuilliers},
  \citenamefont {Bourgeois}, \citenamefont {Kilcher},\ and\ \citenamefont
  {Letessier}}]{Beraud1977}%
  \BibitemOpen
  \bibfield  {author} {\bibinfo {author} {\bibfnamefont {R.}~\bibnamefont
  {B{\'{e}}raud}}, \bibinfo {author} {\bibfnamefont {M.}~\bibnamefont {Meyer}},
  \bibinfo {author} {\bibfnamefont {M.}~\bibnamefont {Desthuilliers}}, \bibinfo
  {author} {\bibfnamefont {C.}~\bibnamefont {Bourgeois}}, \bibinfo {author}
  {\bibfnamefont {P.}~\bibnamefont {Kilcher}}, \ and\ \bibinfo {author}
  {\bibfnamefont {J.}~\bibnamefont {Letessier}},\ }\href {\doibase
  10.1016/0375-9474(77)90119-1} {\bibfield  {journal} {\bibinfo  {journal}
  {Nuclear Physics A}\ }\textbf {\bibinfo {volume} {284}},\ \bibinfo {pages}
  {221} (\bibinfo {year} {1977})}\BibitemShut {NoStop}%
\bibitem [{\citenamefont {Rasmussen}(1959)}]{Rasmussen1959}%
  \BibitemOpen
  \bibfield  {author} {\bibinfo {author} {\bibfnamefont {J.~O.}\ \bibnamefont
  {Rasmussen}},\ }\href {\doibase 10.1103/PhysRev.113.1593} {\bibfield
  {journal} {\bibinfo  {journal} {Physical Review}\ }\textbf {\bibinfo {volume}
  {113}},\ \bibinfo {pages} {1593} (\bibinfo {year} {1959})}\BibitemShut
  {NoStop}%
\bibitem [{\citenamefont {Kondev}\ \emph {et~al.}(2018)\citenamefont {Kondev},
  \citenamefont {Juutinen},\ and\ \citenamefont {Hartley}}]{Kondev2018}%
  \BibitemOpen
  \bibfield  {author} {\bibinfo {author} {\bibfnamefont {F.}~\bibnamefont
  {Kondev}}, \bibinfo {author} {\bibfnamefont {S.}~\bibnamefont {Juutinen}}, \
  and\ \bibinfo {author} {\bibfnamefont {D.}~\bibnamefont {Hartley}},\ }\href
  {\doibase 10.1016/j.nds.2018.05.001} {\bibfield  {journal} {\bibinfo
  {journal} {Nuclear Data Sheets}\ }\textbf {\bibinfo {volume} {150}},\
  \bibinfo {pages} {1} (\bibinfo {year} {2018})}\BibitemShut {NoStop}%
\bibitem [{\citenamefont {Schuessler}\ \emph {et~al.}(1995)\citenamefont
  {Schuessler}, \citenamefont {Benck}, \citenamefont {Buchinger},\ and\
  \citenamefont {Carter}}]{Schuessler1995}%
  \BibitemOpen
  \bibfield  {author} {\bibinfo {author} {\bibfnamefont {H.}~\bibnamefont
  {Schuessler}}, \bibinfo {author} {\bibfnamefont {E.}~\bibnamefont {Benck}},
  \bibinfo {author} {\bibfnamefont {F.}~\bibnamefont {Buchinger}}, \ and\
  \bibinfo {author} {\bibfnamefont {H.}~\bibnamefont {Carter}},\ }\href
  {\doibase 10.1016/0168-9002(95)90009-8} {\bibfield  {journal} {\bibinfo
  {journal} {Nuclear Instruments and Methods in Physics Research Section A:
  Accelerators, Spectrometers, Detectors and Associated Equipment}\ }\textbf
  {\bibinfo {volume} {352}},\ \bibinfo {pages} {583} (\bibinfo {year}
  {1995})}\BibitemShut {NoStop}%
\end{thebibliography}%
\end{document}